\documentclass[aps,onecolumn,floatfix,nofootinbib]{revtex4}
\usepackage{graphicx,epsfig,wrapfig,color}
\usepackage{amsmath}
\usepackage{amssymb}
\usepackage{dsfont}
\usepackage{xcolor}
\usepackage{comment}
\usepackage{color}

\newcommand{\be}{\begin{equation}}
\newcommand{\ee}{\end{equation}}
\newcommand{\ba}{\begin{eqnarray}}
\newcommand{\ea}{\end{eqnarray}}

\newcommand{\la}{\langle}
\newcommand{\ra}{\rangle}

\usepackage{ulem}

\begin{document}
\title{\boldmath
Exactly solvable models of nonlinear extensions of the Schr\"odinger equation}
\author{Tom Dodge$^{1,2}$, Peter Schweitzer$^1$}
\affiliation{
    $^1$
    Physics Department,
    City College of San Francisco,
    San Francisco, CA 94124, U.S.A.\\
    $^2$
    Department of Physics,
    University of Connecticut,
    Storrs, CT 06269-3046, U.S.A.}
 \date{March, 2023}
\begin{abstract}
A method is presented to construct exactly solvable 
nonlinear extensions of the Schr\"odinger equation.
The method explores a correspondence which can be established under
certain conditions between exactly solvable ordinary Schr\"odinger 
equations and exactly solvable nonlinear theories.
We provide several examples illustrating the method. We rederive
well-known soliton solutions and find new exactly solvable nonlinear 
theories in various space dimensions which, to the best of our knowledge, 
have not yet been discussed in literature. Our method can be used to 
construct further nonlinear theories and generalized to relativistic 
soliton theories, and may have many applications. 
\end{abstract}



\maketitle

\section{Introduction}

It is ``quite a rarity in the world of nonlinear differential equations''
to encounter exact analytic solutions \cite{Shifman:2012zz}.
While some exact solutions of nonlinear theories are known, 
see for instance
\cite{tHooft:1974kcl,Polyakov:1974ek,BialynickiBirula:1976zp,
BialynickiBirula:1979dp,Oficjalski:1978,Zakharov-Shabat},
the above quote from Ref.~\cite{Shifman:2012zz} nicely illustrates 
that in general they are rare. 
The goal of this work is to present a method allowing one to construct 
systematically exactly solvable nonlinear theories. We will focus on a 
specific class of nonlinear differential equations, namely on nonlinear 
extensions of the Schr\"odinger equation (NSE). 
The ordinary Schr\"odinger equation (SE) of nonrelativistic quantum 
mechanics is, of course, linear. But its nonlinear extensions have received 
considerable attention in literature and have numerous applications 
in a variety of contexts 
\cite{BialynickiBirula:1976zp,
BialynickiBirula:1979dp,Oficjalski:1978,Zakharov-Shabat,
Rosen:1969ay,Hudson:2017xug,Belendryasova:2021jgs,
Gross,Pitaevskii,Lieb-Liniger:1963,Askaryan,Talanov,Chiao,Kelley:1965,
Tsuzuki:1971,Konotop:2004,Karpiuk:2012,
dark-soliton-a,dark-soliton-b,Yefsah:2013,
Peregrine,rogue-ocean,Wang:2020mjt,rogue-optical,
Guth:2014hsa,Kormos:2010,Beg:1984yh,Namjoo:2017nia,
Weinstein:1983,Alvarez:1997ma,Serkin-Hasegawa:2000,
Blas:2015hro,Biondini:2015pnf,Koch:2022}.

In this work, we will show that under certain circumstances it is possible, 
starting from a known exact analytic ground state solution of a quantum 
mechanical problem, to construct an exactly solvable nonlinear theory. 
We will illustrate the method by providing several examples. In each case, 
the starting point is an exactly solvable quantum problem described by an 
ordinary SE like the harmonic oscillator, Coulomb problem, and other examples. 
As a result, we will derive nonlinear theories which have exact analytic
solutions. In two of the cases, we will rederive well-known soliton solutions.
In several other cases we will present exactly solvable NSEs which have not 
been discussed in literature before to the best of our knowledge. The method 
can be explored to construct systematically further exactly solvable 
nonlinear theories and can be generalized to relativistic theories.

Besides being of immense interest for their own sake, exactly solvable NSEs  
can provide useful toy models and theoretical test grounds in many situations. 
For instance, the availability of exact analytic solutions of nonlinear 
theories can be used to effectively test numerical methods for 
nonlinear partial differential equations.
The numerous applications of NSE theories range from particle physics
\cite{Rosen:1969ay,Hudson:2017xug,Belendryasova:2021jgs}, to many body 
systems and propagation of light through nonlinear media
\cite{Gross,Pitaevskii,Lieb-Liniger:1963,Askaryan,Talanov,Chiao,Kelley:1965,
Tsuzuki:1971,Konotop:2004,Karpiuk:2012,dark-soliton-a,dark-soliton-b,Yefsah:2013}, 
to descriptions of rogue waves in oceans or optics
\cite{Peregrine,rogue-ocean,Wang:2020mjt,rogue-optical},
to cosmological models~\cite{Guth:2014hsa}. 
NSEs emerge naturally in the context of the transition from 
relativistic quantum field theories to nonrelativistic domains
\cite{Kormos:2010,Beg:1984yh,Namjoo:2017nia} 
and play an important role in mathematical physics
\cite{Weinstein:1983,Alvarez:1997ma,Serkin-Hasegawa:2000,
Blas:2015hro,Biondini:2015pnf,Koch:2022}.

Another important application of studies of NSE is to provide 
frameworks for experimental tests of the linearity of quantum mechanics. 
Different schemes have been proposed \cite{Weinberg:1989cm,Kaplan-Rajendran} 
and used to establish upper limits for nonlinear behavior 
in quantum mechanics based on neutron interferometry 
\cite{Shull:1980zz,Gahler:1981zz}, measurements in quantum bound states 
\cite{Bollinger:1989zz,Chupp,Walsworth:1990zz,Majumder:1990zz},
or Ramsey interferometry of vibrational modes of trapped ions 
\cite{Broz:2022aea}. 
So far, no deviations from linear behavior have been observed, and 
it is of importance to establish more stringent experimental limits.

This work is organized as follows. In Sec.~\ref{Sec-2:construction},
we introduce the notation and present the method to construct an
analytically solvable NSE based on an analytically solvable SE.
In Secs.~\ref{Sec-3:Gausson} and \ref{Sec-4:Gausson-trapped}, we 
explore the exactly solvable quantum harmonic oscillator to rederive 
a NSE describing a free or trapped Gausson in any number of space 
dimensions which has been encountered previously, independently 
in different theoretical settings. 
In Sec.~\ref{Sec-5:one-over-cosh}, we will rederive the well-known
one-dimensional $1/\cosh$-soliton and generalize it to any number 
of dimensions in Sec.~\ref{Sec-6:one-over-cosh-N}.
The latter as well as the examples presented subsequently have not been 
discussed in literature before to the best of our knowledge and 
constitute novel results. This includes the exactly solvable NSE with an 
arbitrary power-like nonlinearity in Sec.~\ref{Sec-7:one-over-cosh-variant}
and the NSE derived from a special case of the Rosen-Morse potential 
in Sec.~\ref{Sec-8:tan2}. In Sec.~\ref{Sec-9:tam-1D}, we construct an
interesting NSE based on an exactly solvable potential which contains 
the $\delta$-function potential as a limiting case. Our last example 
is an NSE derived from the exactly solvable Coulomb potential.
Some of these examples are formulated in $N=1$ or $N=3$ dimensions, 
but several of them are formulated for general $N$.
Our conclusions are presented in Sec.~\ref{Sec-11:conclusions}.
The Appendix~\ref{App:A} contains technical details on an interesting
limiting situation.

\newpage

\section{Construction of exactly solvable NSE\lowercase{s}}
\label{Sec-2:construction}

Let us begin with a remark regarding notation. In the NSE literature, often 
a unit system is used with $\hbar=1$ and many authors consider a particle 
of unit mass $m=1$ or set $2m=1$. In this work, we will explicitly use 
SI units and keep all physical constants in the equations. This will
allow the reader to implement her or his preferred notation.

The starting point is ordinary quantum mechanics in $N$ space dimensions
of a nonrelativistic spin-0 particle of mass $m$ moving in a potential 
$U(\vec{x})$ which is described by the linear Schr\"odinger equation (SE)
\be\label{Eq:SE}
      i\hbar \,\frac{\partial\Psi(t,\vec{x})}{\partial t} 
      = -\,\frac{\hbar^2}{2m}\,\bigtriangleup \Psi(t,\vec{x}) 
      + U(\vec{x})\,\Psi(t,\vec{x})\, .
\ee 
We shall assume the potential to be spherically symmetric such that 
$U(\vec{x})=U(r)$ with $r=|\vec{x}|$ for $N\ge 2$ dimensions. For $N=1$,
we shall assume the potential $U(x)$ to be even. The 
$N$-dimensional Laplace operator is given by 
\be\label{Eq:Laplace}
      \bigtriangleup = 
      \frac{1}{r^{N-1}}\,\frac{\partial}{\partial r}\,
      r^{N-1}\,\frac{\partial}{\partial r} + \dots 
      = \frac{\partial^2}{\partial r^2} 
      + \frac{N-1}{r}\,\frac{\partial}{\partial r} + \dots\;\;
\ee
where, for $N\ge 2$, the dots indicate derivatives with respect to angular 
variables which will not be needed because we will focus exclusively on 
ground state wave functions depending solely on $r$ for a spherically 
symmetric potential. The space dimension $N$ will always be clear from 
the context.

Let the potential in  Eq.~(\ref{Eq:SE}) be such that it admits at least one 
bound state. We denote the ground state energy by $E_0$ and the ground state 
wave function by 
\be\label{Eq:soliton-at-rest}
     \Psi_0(t,\vec{x}) = c_0\,\phi_0(r)\,e^{-iE_0t/\hbar}
\ee
with the normalization $\int d^Nr\,| \Psi_0(\vec{x},t)|^2=1$.
Due to the symmetry of the potential, the spatial part of 
$\Psi_0(t,\vec{x})$ is described by a radial function $\phi_0(r)$ 
for $N\ge 2$ space dimensions. For $N=1$, the wave function $\phi_0(x)$ 
is even. For the following, it will be convenient to choose the phase and 
define the normalization constant $c_0>0$ in Eq.~(\ref{Eq:soliton-at-rest}) 
such that
\be\label{Eq:norm-phi-0}
      0 \le \phi_0(r) \le 1\,, \quad  \phi_0(0)=1\,.
\ee

After these preparations, we are in the position to present the method.
If the quantum mechanical problem in Eq.~(\ref{Eq:SE}) can be solved 
analytically, then, depending on the properties of the radial function
$\phi_0(r)$, it may be possible to invert $\phi_0(r)$ and find a 
function $F$ such that the potential can be expressed as
\be\label{Eq:pot-vs-nonlin}
       U(r) = F[\Psi^\ast\Psi]\biggl|_{\Psi = \Psi_0(t,\vec{x})} \,.
\ee
If this step can be carried out, then $F$ will in general be a nonlinear 
function of the wave function $\Psi$. This allows us to rewrite the SE in 
Eq.~(\ref{Eq:SE}) in terms of a nonlinear extension of the Schr\"odinger 
equation (NSE) as follows
\be\label{Eq:NSE}
      i\hbar \,\frac{\partial\Psi}{\partial t} 
      = -\,\frac{\hbar^2}{2m}\,\bigtriangleup\Psi 
      + F[\Psi^\ast\Psi]\,\Psi\,.
\ee 
Notice that it is convenient to choose $\Psi^\ast\Psi$ as variable of the
nonlinear function in Eq.~(\ref{Eq:pot-vs-nonlin}) because in this way 
the NSE (\ref{Eq:NSE}) is linear with respect to the phase of $\Psi$
which carries the information about the time dependence.

The NSE (\ref{Eq:NSE}) has the exact, analytically known solution given by 
Eq.~(\ref{Eq:soliton-at-rest}) which corresponds to a stationary soliton 
solution in the corresponding nonrelativistic nonlinear theory. A soliton 
traveling with a constant velocity $\vec{v}$ can be obtained by applying 
a Galilean boost to Eq.~(\ref{Eq:soliton-at-rest}) as follows
\be\label{Eq:soliton-in-motion}
      \Psi(t,\vec{x})=c_0\,\phi_0(\vec{x}-\vec{v}t)\,
      e^{i(m\vec{v}\cdot\vec{x}-\frac12m\vec{v}^2t-E_0t)/\hbar}\,.
\ee
The crucial step in this construction is the derivation of the function 
$F[\Psi^\ast\Psi]$. For a spherically symmetric potential $U(\vec{x})=U(r)$ 
in $N\ge2$ (or even potential $U(x)$ in $N=1$), it may be possible to carry 
out this step if $\phi_0(r)$ is monotonously decreasing and an inverse 
function $\phi_0^{-1}$ exists such that $\phi_0^{-1}[\phi_0^{ }(r)]=r$ 
(analogously for $N=1$). 
In our context, it will be important that this crucial step can be carried 
out {\it analytically} which ultimately depends on the properties of the 
potential.

In the following sections, we will discuss examples to illustrate the method. 
Hereby, we will focus on the construction of exactly solvable NSEs with 
analytic solutions. Such exactly solvable nonlinear theories are of interest 
for their own sake and may have interesting applications. 
In principle, further work is required to establish that a solution of a
NSE of the type (\ref{Eq:soliton-in-motion}) can be considered a soliton
in the strict mathematical sense. For that it would be important to 
show, for instance, that two such solutions can scatter off each other 
and will preserve their shapes long before and long after the scattering 
process. Such investigations are beyond the scope of this work, but have 
been carried out in literature in some cases and we shall refer to
them in the following.

\newpage
\section{\boldmath $N$-dimensional logarithmic nonlinear theory, 
the free Gausson} 
\label{Sec-3:Gausson}

As a first example, we consider the harmonic oscillator in 
$N$-dimensional space. The system is defined by the SE in 
Eq.~(\ref{Eq:SE}) with a harmonic potential 
\be\label{Eq:harm-osc-pot}
        U(r) = \frac12\,m\,\Omega^2\,r^2 \,.
\ee
The ground state energy and wave function are given by
\be\label{Eq:harm-osc-WF}
    \Psi_0(t,\vec{x})=c_0\,\phi_0(r)\,e^{-iE_0t/\hbar}, 
    \quad \phi_0(r) = e^{-r^2/b^2},
    \quad b = \sqrt{\frac{2\hbar}{m\Omega}}\,,
    \quad c_0 = \biggl(\frac{m\Omega}{\pi\hbar }\biggr)^{\!N/4}\,,
    \quad E_0 = \frac12\,N\,\hbar\Omega\,.
\ee
Inverting the wave function as 
\be\label{Eq:harm-osc-rewrite-1}
    r^2 = - \frac{\hbar}{m\Omega}\;\ln
    \biggl(\frac{|\Psi_0(t,\vec{x})|^2}{|c_0|^2}\biggr)\,\
\ee
we can rewrite the harmonic potential as
\be\label{Eq:harm-osc-rewrite-2}
    U(r)= \frac12\,m\Omega^2 r^2 
        = -\frac{\hbar\Omega}{2}\;\ln
        \biggl(\frac{|\Psi_0(t,\vec{x})|^2}{|c_0|^2}\biggr) \,.
\ee
In this way, we derive the NSE (\ref{Eq:NSE}) with a logarithmic 
nonlinear term $F[\Psi^\ast\Psi]$ defined as follows
\be\label{Eq:harm-osc-NSE-2}
      F\bigl[\Psi^\ast\Psi\bigr]
      = - A \,\ln\bigl(B \,|\Psi|^2\bigr)\,,
      \quad A = \frac{\hbar\Omega}{2},
      \quad B = \frac{1}{|c_0|^2}\,.
\ee
The exactly solvable NSE in 
Eqs.~(\ref{Eq:NSE},~\ref{Eq:harm-osc-NSE-2}) with the analytic solution 
(\ref{Eq:harm-osc-WF}) is known as the nonrelativistic Gausson, and was 
studied in detail in 
\cite{BialynickiBirula:1979dp,BialynickiBirula:1976zp,Oficjalski:1978} 
including relativistic formulations. Previously, these solutions were
encountered in $N=3$ dimensions in studies of relativistic theories 
invariant under space-time dilatations \cite{Rosen:1969ay}. Much later, 
Gaussons were rediscovered in a study of the energy-momentum tensor 
where point-like particles were ``smeared out'' to simulate an internal 
structure \cite{Hudson:2017xug}. Recently, relativistic one-dimensional 
Gaussons were studied in \cite{Belendryasova:2021jgs}.

\begin{figure}[b!] 
\includegraphics[width=0.32\linewidth]{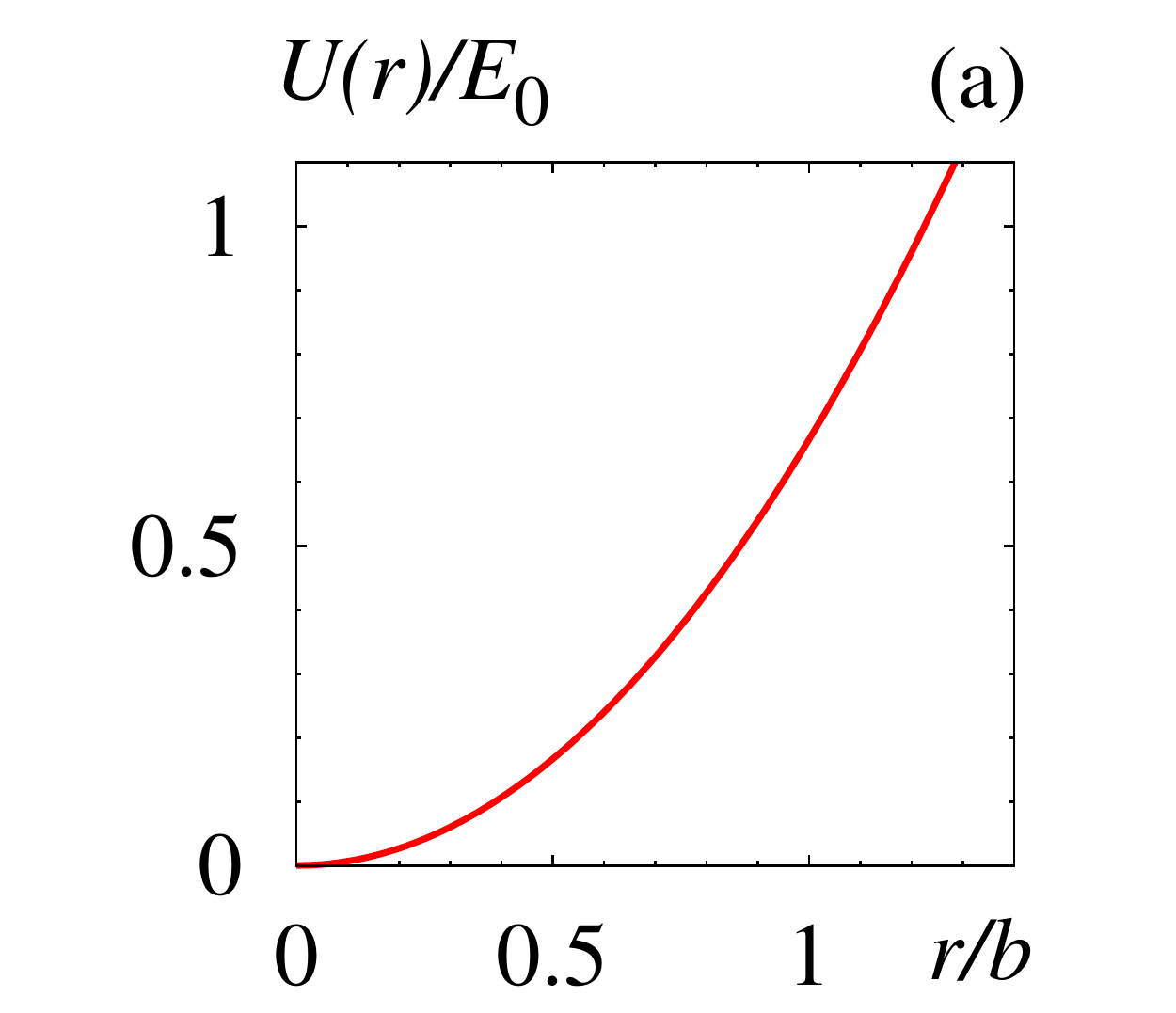}
\includegraphics[width=0.32\linewidth]{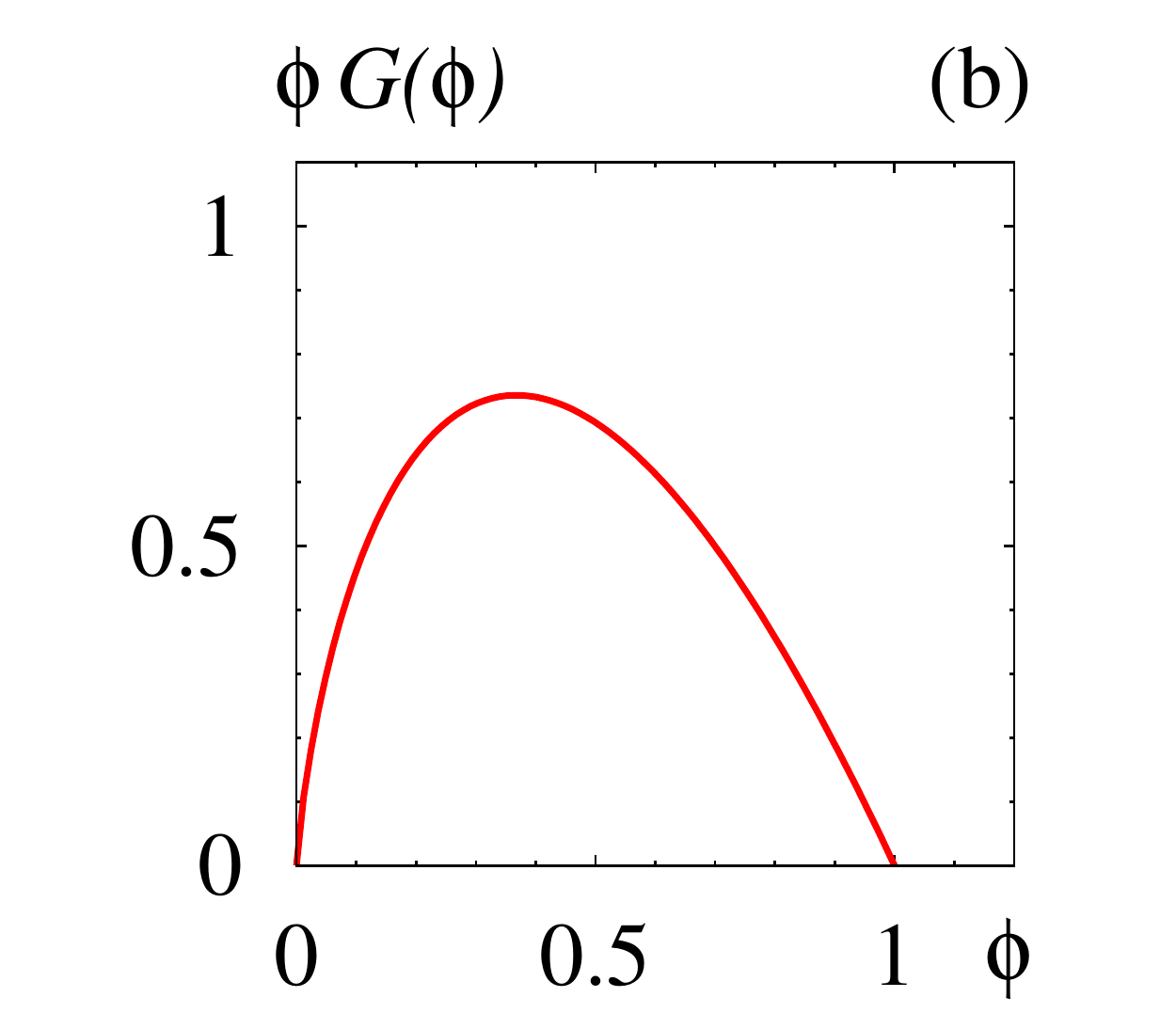}
\includegraphics[width=0.32\linewidth]{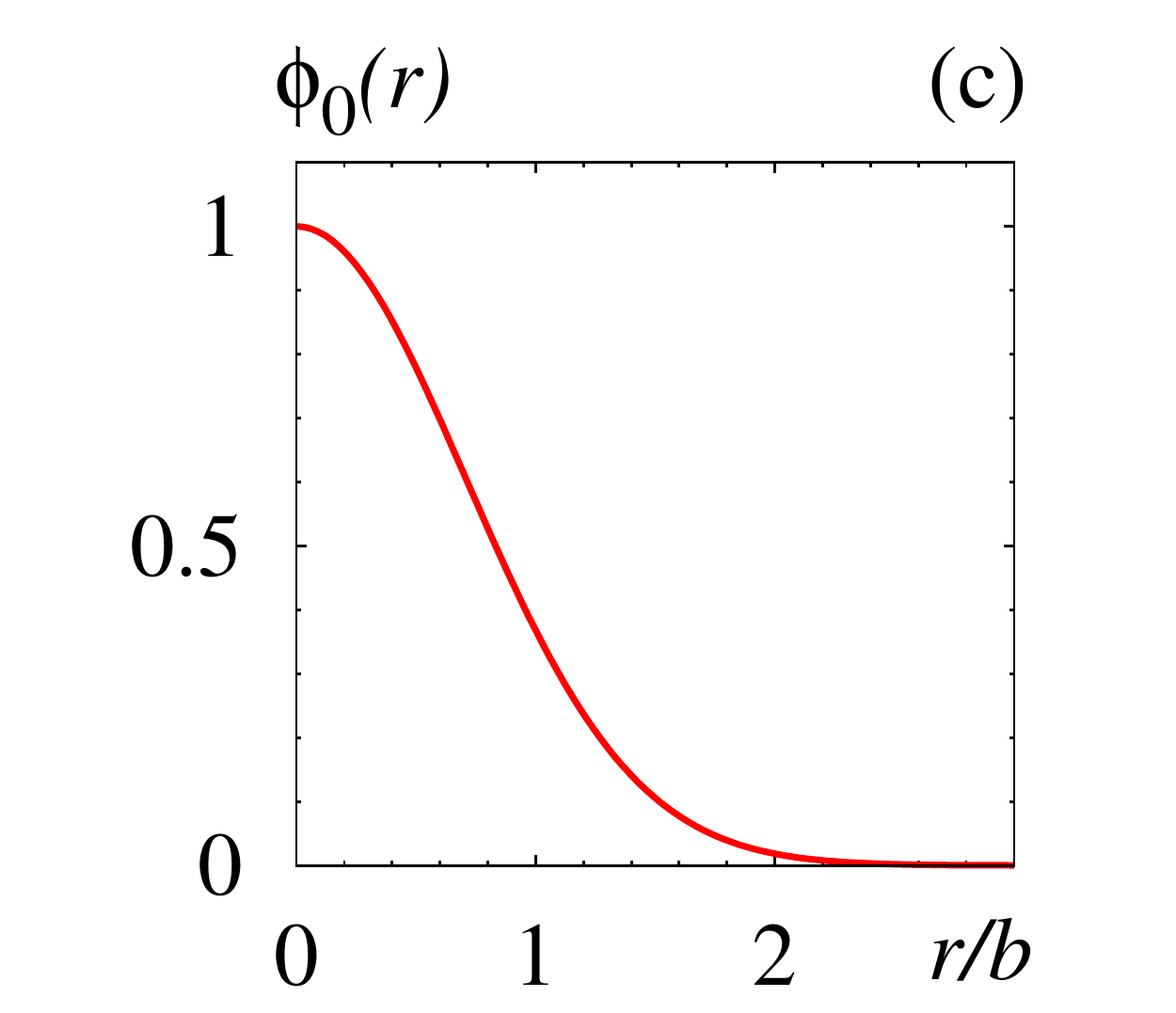}
\caption{(a) The harmonic oscillator potential $U(r)$ in units 
of ground state energy $E_0$ as a function of $r$ in units of 
$b=\sqrt{2\hbar/(m\Omega)}$ in the 3-dimensional case.
(b) The nonlinear term $\phi\, G(\phi)$ of the NSE 
defined in Eq.~(\ref{Eq:harm-osc-NSE-3}). 
(c) The radial part $\phi(r)$ of the ground state wave function 
of the SE with the potential in (a) and the soliton of the 
NSE defined by the nonlinear term in (b).
\label{Fig-1:harm-osc}}
\end{figure}

In Fig.~\ref{Fig-1:harm-osc}, we show the potential of the SE, 
the nonlinear term of the NSE, and the radial part of the wave function
(the potential and $\phi_0(r)$ of the harmonic oscillator are well known, but we 
include them for completeness and consistency with the following sections). 
It is convenient to display the potential in Fig.~\ref{Fig-1:harm-osc}a in 
units of the ground state energy and $r$ in units of 
$b=\sqrt{2\hbar/(m\Omega)}$. 
The nonlinear term (\ref{Eq:harm-osc-NSE-2}) is visualized as a
function of the dimensionless variable $\phi$ as 
\be\label{Eq:harm-osc-NSE-3}
      F\bigl[\Psi^\ast\Psi\bigr]  = A \, G\bigl(|\Psi|/c_0\bigr)\,,
      \quad G(\phi) = - {\rm ln}\,\phi^2 .
\ee
Recalling the normalization and phase convention in Eq.~(\ref{Eq:norm-phi-0}),
the variable $\phi$ satisfies $0\le \phi \le 1$. For $ 0 < \phi < 1$ the 
function $G(\phi)$ is positive. As $\phi\to0$ the nonlinear term $G(\phi)$ 
diverges which reflects the growth of $U(r)$ for $r\to\infty$. However, 
the nonlinearity enters in Eqs.~(\ref{Eq:NSE},~\ref{Eq:harm-osc-NSE-2}) 
practically as $\phi\,G(\phi)$ which goes to zero for $\phi\to0$ and $\phi\to1$ 
assuming its maximum in between at $\phi=1/e$, i.e. at this point the 
nonlinearity in Eqs.~(\ref{Eq:NSE},~\ref{Eq:harm-osc-NSE-2}) 
is strongest, see Fig.~\ref{Fig-1:harm-osc}b.
The radial part $\phi_0(r)$ of respectively the ground state 
wave function of the ordinary SE and the soliton of the NSE 
is shown in Fig.~\ref{Fig-1:harm-osc}c.

The solution (\ref{Eq:harm-osc-WF}) corresponds to a Gausson at rest.
By applying the Galilean boost in Eq.~(\ref{Eq:soliton-in-motion}) to 
the solution (\ref{Eq:harm-osc-WF}), we obtain a soliton traveling 
with constant velocity which preserves its shape. In our derivation, 
the shape-preserving traveling solution appears as a trivial 
consequence of Galilean invariance of the NSE in
Eqs.~(\ref{Eq:NSE},~\ref{Eq:harm-osc-NSE-2}).
As remarked in Sec.~\ref{Sec-2:construction}, dedicated analyses are needed 
to show that such shape-preserving solutions can scatter off each other and 
asymptotically (i.e.\ long before and long after the scattering event) 
preserve their shapes. In the case of the Gausson solution,this was 
shown in \cite{BialynickiBirula:1979dp,BialynickiBirula:1976zp}.
Noteworthy is the existence of a ``resonance region'' in which the 
scattering can be inelastic and the collision of two Gaussons can
produce a final state with three Gaussons \cite{Oficjalski:1978}.

\

\

\section{Theory of a Gausson trapped in a harmonic potential} 
\label{Sec-4:Gausson-trapped}

The steps carried out in Sec.~\ref{Sec-3:Gausson} can be performed
also for a ``part''  of the potential leading to the NSE of a Gausson
trapped in the ``remaining part'' of the harmonic potential. 
For definiteness, we will consider $N=3$ space dimensions, but a 
generalization to other space dimensions $N$ is straightforward 
and analogous to Sec.~\ref{Sec-3:Gausson}. 

For that, let us consider the harmonic potential 
$U(r) = \frac12 \,m\,\Omega_1^2 \,r^2 + \frac12\, m\,\Omega_2^2 \,r^2$ 
with $\Omega_1^2+\Omega_2^2=\Omega^2$.
Now we choose the part $\frac12\,m\,\Omega_1^2 \, r^2$ of the potential 
to be left alone and reformulate the part $\frac12\,m\,\Omega_2^2\,r^2$ 
in terms of the nonlinear theory as discussed in Sec.~\ref{Sec-3:Gausson}. 
In this way, we obtain the following NSE
\be\label{Eq:harm-osc-NSE-trapped-1}
      i\hbar \,\frac{\partial\Psi}{\partial t} 
      = -\,\frac{\hbar^2}{2m}\,\bigtriangleup\Psi + U_1(r) \, \Psi
      + F_2\bigl[\Psi^\ast\Psi] \;\Psi \,,
\ee   
where the potential $U_1(r)$, the nonlinear term $F_2\bigl[\Psi^\ast\Psi]$
and the constants $A_2>0$ and $B_2>0$ are given by 
\be\label{Eq:harm-osc-NSE-trapped-2}
      U_1(r) = \frac12 \,m\,\Omega_1^2 \,r^2, 
      \quad  F_2\bigl[\Psi^\ast\Psi] 
      = - A_2 \,\ln\bigl(B_2 \,\Psi^\ast\Psi\bigr)\,,
      \quad A_2 = \frac{\hbar\Omega_2^2}{2\Omega},
      \quad B_2 = \biggl(\frac{\pi\hbar}{m\Omega}\biggr)^{3/2} \,.
\ee 
Let us recall that these results are specifically for $N=3$ space dimensions
and the generalization to other dimensions is straight forward.
The NSE given by Eqs.~(\ref{Eq:harm-osc-NSE-trapped-1},
\ref{Eq:harm-osc-NSE-trapped-2}) describes a Gausson trapped 
in the harmonic potential $U_1(r) = \frac12 \,m\,\Omega_1^2 \,r^2$ 
with the analytic solution given by Eq.~(\ref{Eq:harm-osc-WF}).

In the limit $\Omega_1 \rightarrow \Omega$ and $\Omega_2 \rightarrow 0$,
the nonlinear theory (\ref{Eq:harm-osc-NSE-trapped-1},
\ref{Eq:harm-osc-NSE-trapped-2}) reduces to the
regular SE for a harmonic oscillator.
In the limit $\Omega_1 \rightarrow 0$ and $\Omega_2 \rightarrow \Omega$,
it reduces to the nonlinear theory of a free Gausson discussed
in Sec.~\ref{Sec-3:Gausson}.

\newpage
\section{\boldmath Nonlinear theory with a 1/cosh soliton in one-dimension}
\label{Sec-5:one-over-cosh}

In our next example, we consider a one-dimensional quantum system described 
by the potential 
\be\label{Sec:cosh-a}
      U(x) = -\,\frac{\hbar^2}{m\,a^2} \frac{1}{\cosh^2{(x/a)}} \,,
\ee 
where $a$ is a positive constant with the dimension of length. 
The ground state solution of the SE reads
\be\label{Sec:cosh-b}
      \Psi_0(t,x) = c_0\,\phi_0(x)\,e^{-iE_0t/\hbar}, 
      \quad \phi_0(x) = \frac{1}{\cosh(x/a)}\,,
      \quad c_0 = \frac{1}{\sqrt{2a}} \,,
      \quad E_0 = -\frac{\hbar^2}{2m\,a^2} \,.
\ee 
Using the method described in Sec.~\ref{Sec-2:construction}, 
the wave function can be inverted and used to express the 
potential in terms of the ground state wave function as follows
\be\label{Sec:cosh-c}
      U(x) = - \frac{2\hbar^2}{ma} \Psi^\ast_0(t,x)\Psi_0^{ }(t,x) \,.
\ee 
The resulting analytically solvable NSE is then given by 
Eq.~(\ref{Eq:NSE}) with a particularly simple nonlinear term
\be\label{Sec:cosh-e}
      F\bigl[\Psi^\ast\Psi\bigr]  =  - A \,|\Psi|^2\,,
      \quad A = \frac{2\hbar^2}{m a} \,.
\ee
The analytic solution (\ref{Sec:cosh-b}) of the nonlinear theory
(\ref{Eq:NSE},~\ref{Sec:cosh-e}) is well known and was found in 
Ref.~\cite{Zakharov-Shabat}. The underlying NSE in $N=3$ is generally 
known as the Gross-Pitaevskii equation \cite{Gross,Pitaevskii} and has 
important applications to the description of interacting Bose gases. 
In $N=1$ dimensions, it is often referred to as the Lieb-Liniger model 
\cite{Lieb-Liniger:1963}. 
The wide range of applications of this nonlinear theory includes
propagation of self-focusing laser beams in nonlinear media 
\cite{Askaryan,Talanov,Chiao,Kelley:1965}, solitons in Bose condensates 
\cite{Tsuzuki:1971,Konotop:2004,Karpiuk:2012,dark-soliton-a,dark-soliton-b}
and fermionic superfluids \cite{Yefsah:2013}, generation of
ocean \cite{Peregrine,rogue-ocean,Wang:2020mjt} and optical 
\cite{rogue-optical} rogue waves, or cosmological axion 
models of nonrelativistic dark matter \cite{Guth:2014hsa}.
The NSE can be derived from, e.g., the nonrelativistic limit 
of the one-dimensional sinh-Gordon model \cite{Kormos:2010}, or 
the complex $\Phi^4$ theory \cite{Beg:1984yh,Namjoo:2017nia}. 
Suffice to say that this NSE is of great interest in mathematical physics 
\cite{Weinstein:1983,Alvarez:1997ma,Serkin-Hasegawa:2000,Blas:2015hro,
Biondini:2015pnf,Koch:2022}.
In the case of solitons in Bose condensates, the sign of the nonlinearity 
is opposite to our result and the equation describes a ``dark soliton'' 
which corresponds to a depletion in the density in the Bose condensate
\cite{Tsuzuki:1971,Konotop:2004,Karpiuk:2012,dark-soliton-a,dark-soliton-b}.
 
For completeness, we remark that the potential $U(x)$ in 
Eq.~(\ref{Sec:cosh-a}) is a special case of the Rosen-Morse potential 
\cite{Rosen-Morse:1932} and belongs to a wider class of potentials known 
as Natanzon potentials \cite{Natanzon:1979}. 
We postpone displaying the potential, nonlinear term, and wave
function to the next section where we generalize the 1/cosh solution
to an arbitrary number of dimensions $N$.

\newpage
\section{\boldmath Nonlinear theory with a 1/cosh soliton in $N$ dimensions}
\label{Sec-6:one-over-cosh-N}

The 1/cosh solitons exist also in $N>1$ dimensions, albeit the starting point 
is then a somewhat more complicated quantum mechanical potential which contains
an additional term proportional to $(N-1)$ and is given by
\be\label{Sec:N-DIM-cosh-a}
      U(r) = 
      - \,\frac{\hbar^2}{ma^2} \frac{1}{\cosh^2{(r/a)}} 
      - \,(N-1)\,\frac{\hbar^2}{2ma\, r} \,\tanh(r/a)
\ee 
where $a$ is a positive constant with the dimension of length. Clearly, for 
$N=1$ we recover the potential of Sec.~\ref{Sec-5:one-over-cosh}. The ground 
state solution of the SE is given by
\be\label{Sec:N-DIM-cosh-b}
      \Psi_0(t,\vec{x})=c_0(N)\,\phi_0(r)\,e^{-iE_0t/\hbar}, 
      \quad \phi_0(r) = \frac{1}{\cosh(r/a)}\,,
      \quad E_0 = -\frac{\hbar^2}{2m\,a^2} \,,
\ee 
and is exactly the same as in Sec.~\ref{Sec-5:one-over-cosh}
except the normalization constant is now given by
\be\label{Sec:N-DIM-cosh-norm}
      c_0(N) = \sqrt{
        \frac{4^{N-1}\,\Gamma(N/2+1)}
             {(2^N-4)\,N\,(N-1)!\,\pi^{N/2} \zeta(N-1)a^N}}\,,
      \quad N > 2\,.
\ee
In the case $N=2$, care is needed because the factor $(2^N-4)$ 
goes to zero while the $\zeta$-function $\zeta(N-1)$ diverges, 
but the product of these factors $(2^N-4)\,\zeta(N-1)\to4\ln2$ 
is finite such that $c_0(2)=1/(a\sqrt{2\pi\ln2})$. For $N=1$ 
the formula (\ref{Sec:N-DIM-cosh-norm}) reproduces the 
normalization constant quoted in Sec.~\ref{Sec-5:one-over-cosh} 
in Eq.~(\ref{Sec:cosh-b}).

Inverting the wave function, the potential can be rewritten as
\be\label{Sec:N-DIM-cosh-c}
      U(r) = A\, G\bigl(|\Psi_0^{ }(t,\vec{x})|/c_0|\bigr), 
      \quad A = \frac{\hbar^2}{ma}\,,
      \quad G(\phi) = -\phi^2-\frac12\,(N-1)
      \frac{\sqrt{1-\phi^2}}{\ln\bigl(1+\sqrt{1-\phi^2}\bigr)-\ln \phi}\,.
\ee 
The resulting analytically solvable NSE is then given by
Eq.~(\ref{Eq:NSE}) with the nonlinear function  
\be\label{Sec:N-DIM-cosh-d}
      F[\Psi^\ast\Psi] = A\:G\bigl(|\Psi/c_0^{ }|\bigr) 
\ee 
with $G(\phi)$ and $A$ defined in Eq.~(\ref{Sec:N-DIM-cosh-c}).
The results are valid for any dimension $N$ including the 
one-dimensional case discussed in Sec.~\ref{Sec-5:one-over-cosh}.
To the best of our knowledge, the solution for $N>1$ has not been
discussed before in literature.

\begin{figure}[b!] 
\includegraphics[width=0.32\linewidth]{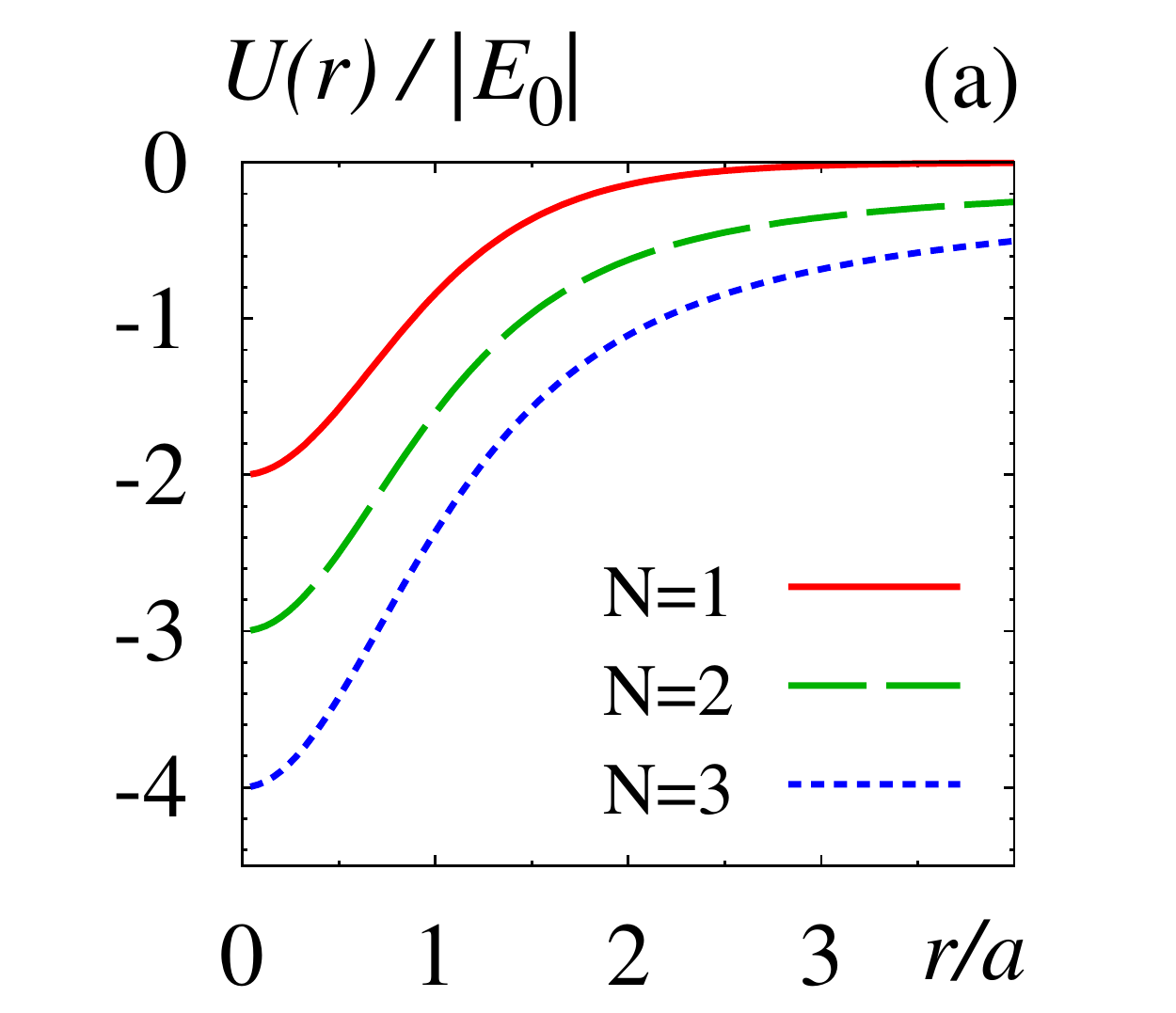}
\includegraphics[width=0.32\linewidth]{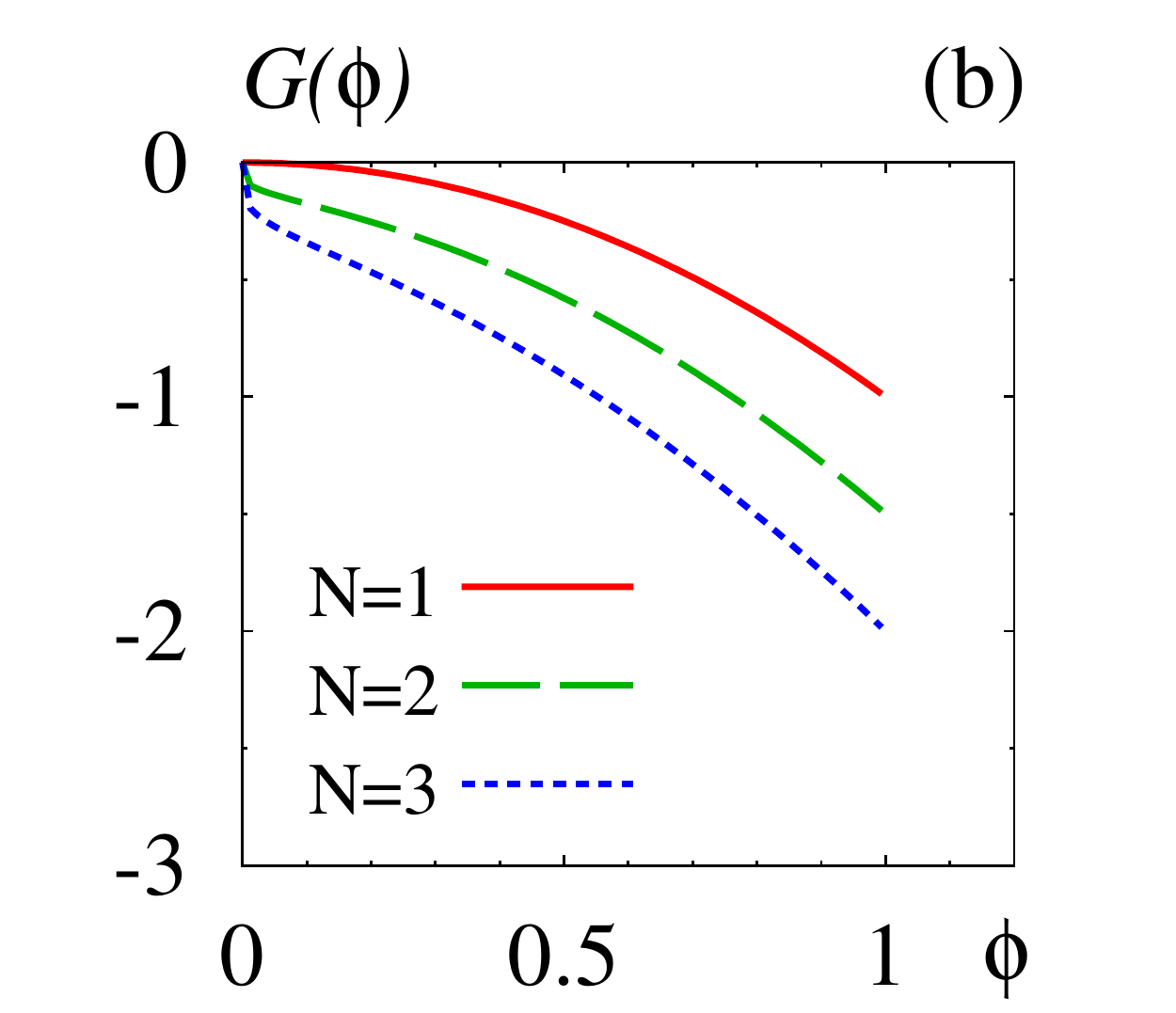}
\includegraphics[width=0.32\linewidth]{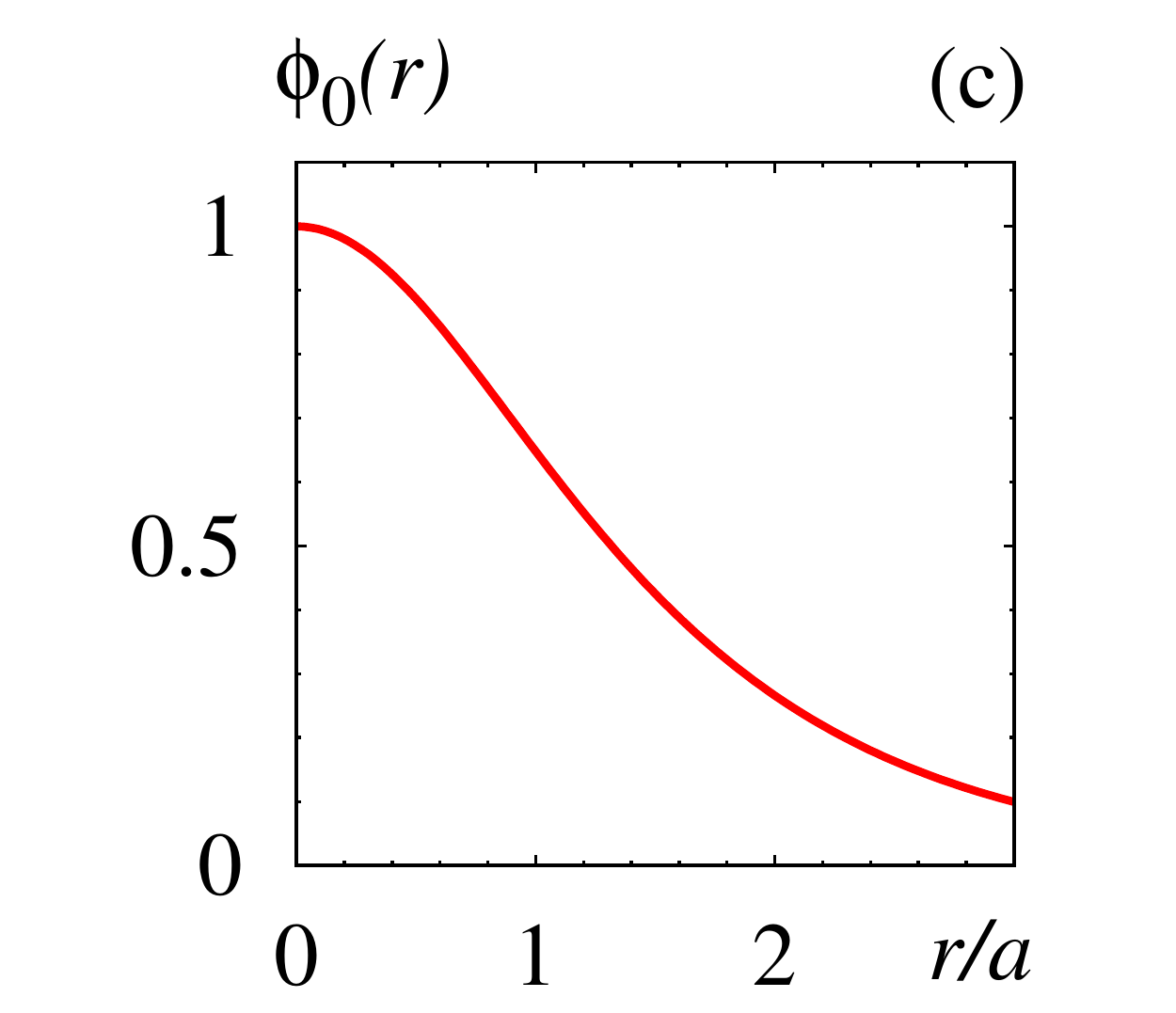}
\caption{(a) The potential $U(r)$ in Eq.~(\ref{Sec:N-DIM-cosh-a})
in units of $|E_0|$ for $N=1,\,2,\,3$ dimensions.
(b) The nonlinear term $G(\phi)$ defined in 
Eq.~(\ref{Sec:N-DIM-cosh-c}) for $N=1,\,2,\,3$ dimensions.
(c) The radial part $\phi_0(r)=1/\cosh(r/a)$ of the ground 
state wave function which is the solution to the SE with
the potential shown in (a) and the NSE with the nonlinear
term shown in (b) for any dimension $N$.
\label{Fig-3:cosh}}
\end{figure}

In units of $|E_0|$, the potential has the shape
$V(r)/|E_0|=-2/\cosh(y)^2-(N-1)\tanh(y)/y$ where $y=r/a$ and is
depicted in Fig.~\ref{Fig-3:cosh}a for $N=1,\,2,\,3$ dimensions.
The function $G(\phi)$ defined in Eq.~(\ref{Sec:N-DIM-cosh-c}) is 
similarly shown for $N=1,\,2,\,3$ dimensions in Fig.~\ref{Fig-3:cosh}b.
Although the nonlinearity in (\ref{Sec:N-DIM-cosh-c}) enters effectively 
as $\phi\, G(\phi)$, we merely plot $G(\phi)$ since in this case the 
nonlinear function vanishes for $\phi\to0$ (in contrast to 
the nonlinearity in the Gausson case in Fig.~\ref{Fig-1:harm-osc}b).
In the limit $\phi\to 1$, the function $G(\phi)$ approaches the value 
$(-N-1)/2$. We see that the nonlinearity in this NSE has a very different 
shape and opposite sign compared to the nonlinearity of the Gausson 
discussed in Secs.~\ref{Sec-3:Gausson} and \ref{Sec-4:Gausson-trapped}.
The radial wave function has the same 1/cosh shape in any dimension and
is shown in Fig.~\ref{Fig-3:cosh}c.

\newpage
\section{\boldmath 
One-dimensional theory with a power-law nonlinearity 
$F[\Psi^\ast_{ }\Psi]=|\Psi^\ast_{ }\Psi|^\lambda$}
\label{Sec-7:one-over-cosh-variant}

In this section, we present an interesting variant of the NSE discussed 
in Sec.~\ref{Sec-5:one-over-cosh}. In a one-dimensional quantum 
system, we consider the potential 
\be\label{Sec:cosh-variant-a}
      U(x) = -\,\frac{1+\lambda}{2\lambda^2}\,
      \frac{\hbar^2}{m\,a^2} \frac{1}{\cosh^2{(x/a)}} \,.
\ee 
where $a>0$ has the dimension of length and $\lambda>0$ is dimensionless.
The case $\lambda = 1$ was discussed in Sec.~\ref{Sec-5:one-over-cosh}, and
for $\lambda\to\infty$ we recover the free SE. For $0<\lambda<\infty$, the
ground state solution of the SE with the potential (\ref{Sec:cosh-variant-a}) 
reads
\be\label{Sec:cosh-variant-b}
      \Psi_0(t,x) = c_0\,\phi_0(x)\,e^{-iE_0t/\hbar}, 
      \quad \phi_0(x) = \frac{1}{\cosh(x/a)^{1/\lambda}}\,,
      \quad c_0 = \sqrt{
        \frac{\Gamma(1/2+1/\lambda)}
             {\sqrt{\pi}\,\Gamma(1/\lambda)\,a}}\,,
      \quad E_0 = -\frac{1}{\lambda^2}\,\frac{\hbar^2}{2m\,a^2} \,.
\ee 
Using the method described in Sec.~\ref{Sec-2:construction}, 
the wave function can be inverted and used to express the 
potential in terms of the ground state wave function as follows
\be\label{Sec:cosh-variant-c}
      U(x) = - \,\frac{1+\lambda}{2\lambda^2}\,
      \frac{\hbar^2}{m\,a^2} \,\bigl(\phi_0^\ast(x)\phi_0^{ }(x)\bigr)^\lambda\,
      \,.
\ee 
The resulting analytically solvable  NSE is then given by
Eq.~(\ref{Eq:NSE}) with with the power-law nonlinear term
\be\label{Sec:cosh-variant-e}
      F\bigl[\Psi^\ast\Psi\bigr]  =  A \,|\Psi|^{2\lambda}\,,
      \quad A =  - \,\frac{1+\lambda}{2\lambda^2}\,
      \frac{2\hbar^2}{m\,a\,c_0^{2\lambda}} \,.
\ee
The constant $\lambda$ can be chosen to model any desired
power-law nonlinearity proportional to $|\Psi|^{2\lambda}$.

\begin{figure}[b!] 
\includegraphics[width=0.32\linewidth]{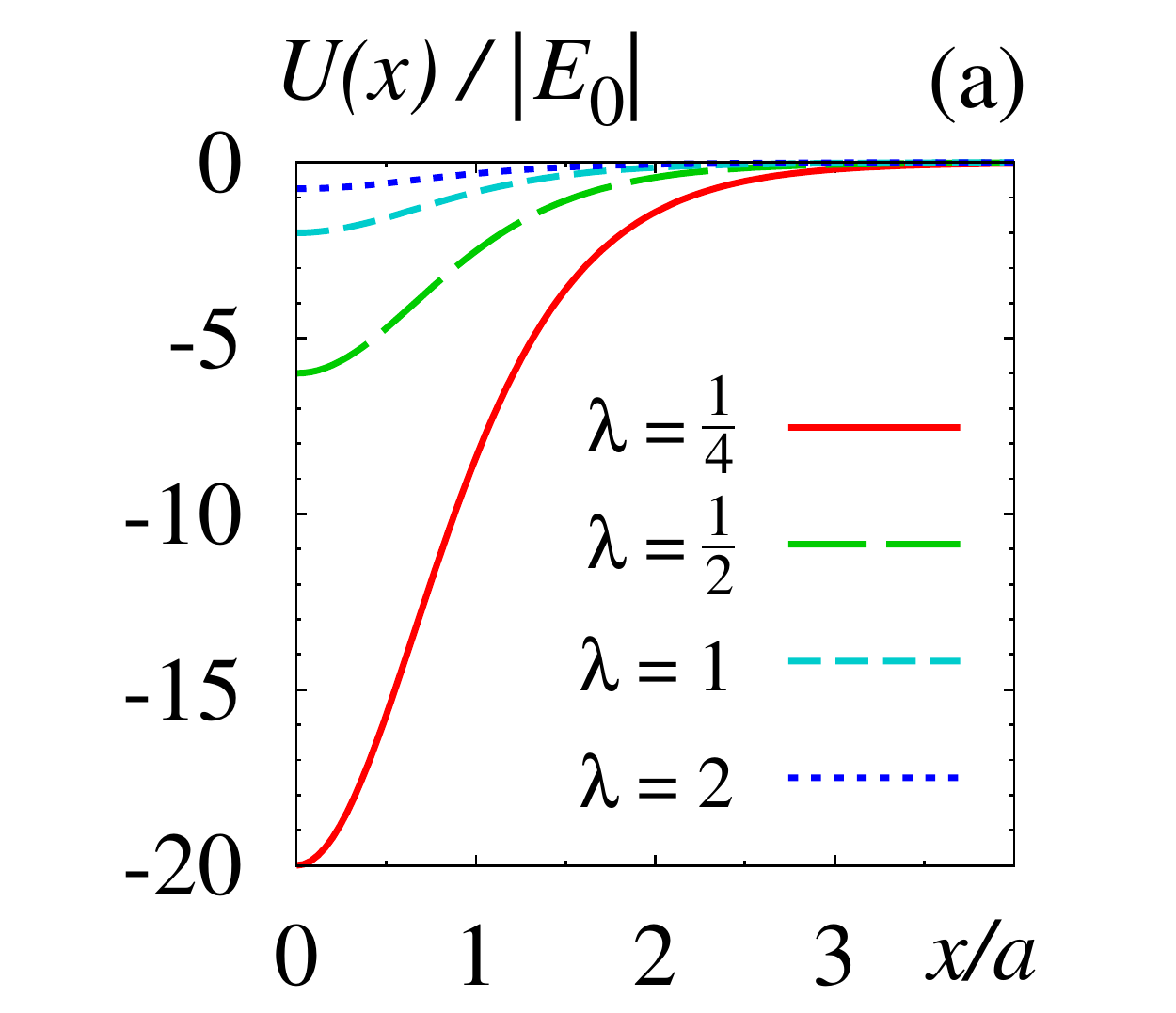}
\includegraphics[width=0.32\linewidth]{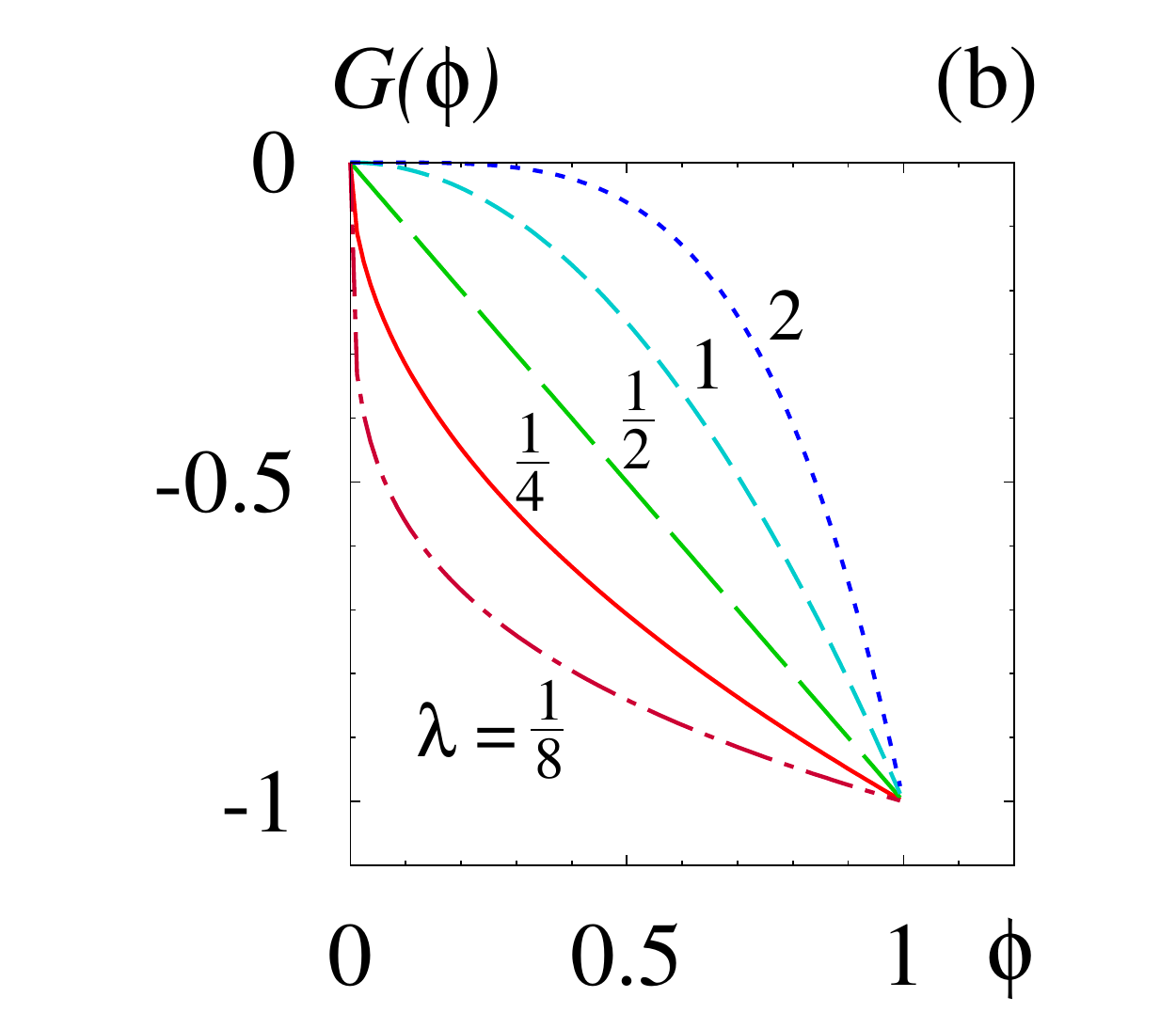}
\includegraphics[width=0.32\linewidth]{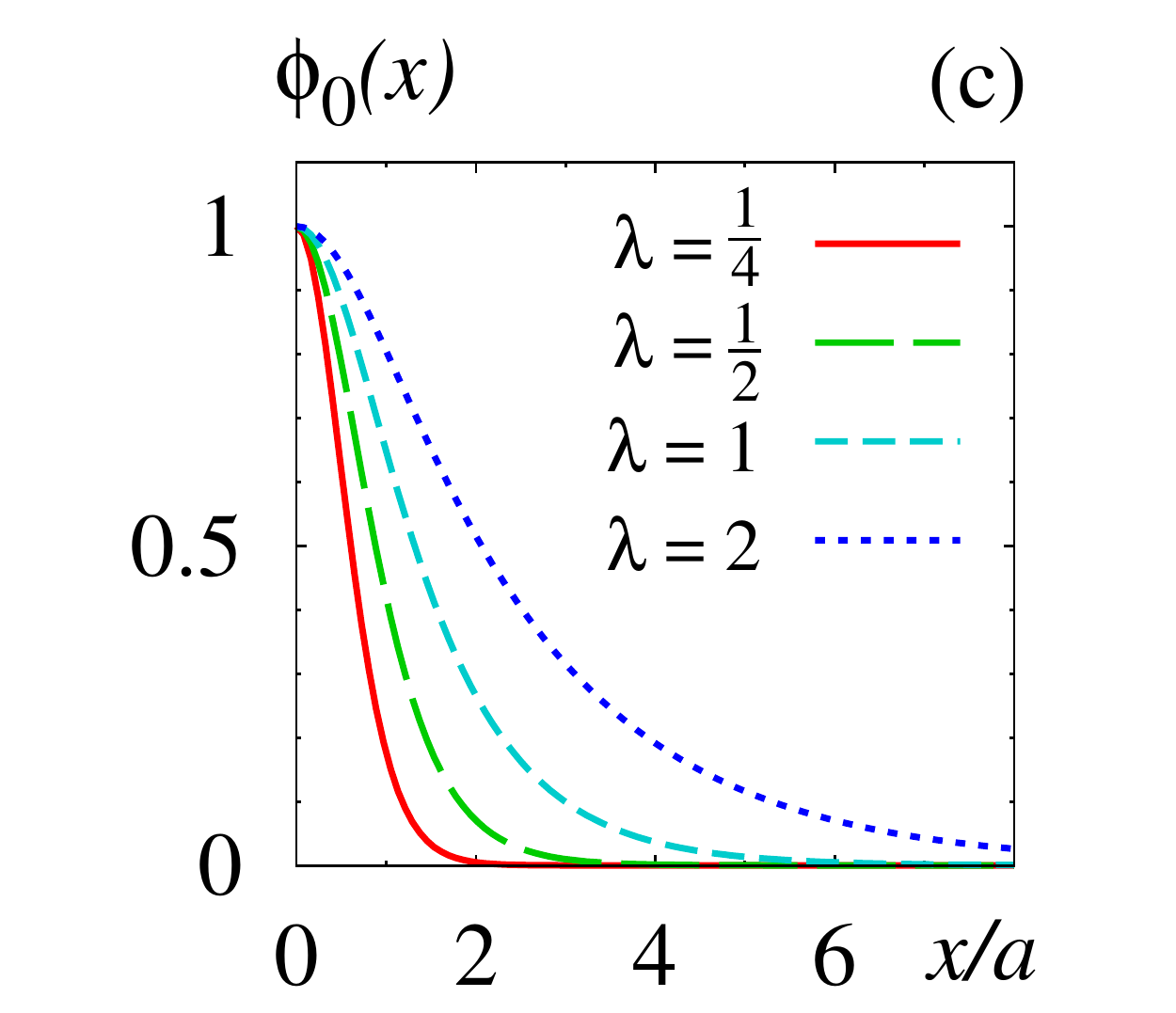}
\caption{
(a) The potential (\ref{Sec:cosh-variant-a}) in units of $|E_0|$, 
(b) the nonlinear term $G(\phi)=-\,\phi^{2\lambda}$ defined 
below Eq.~(\ref{Sec:cosh-variant-e}), and 
(c) the solution $\phi_0(x)=1/\cosh(x/a)^{1/\lambda}$ of the SE with
the potential in (a) and the NSE with the nonlinear term in (b) 
for different $0<\lambda\le 2$.
\label{Fig-4:cosh-lambda}}
\end{figure}

In Fig.~\ref{Fig-4:cosh-lambda}, we show the potential $U(x)$, 
the nonlinear term defined as $G(\phi)=-\,\phi^{2\lambda}$, and 
$\phi_0(x)$ for selected $\lambda$ values.
As $\lambda$ increases, $U(x)$ becomes shallower and $G(\phi)$
more strongly peaked towards the region $\phi\to1$.
In absolute units, the spatial part of the soliton is 
$c_0\,\phi_0(x)$ and the normalization constant $c_0$ decreases
as $\lambda$ increases. I.e., in the limit when $\lambda$ becomes 
large, the soliton decreases in the center and spreads out,
i.e.\ it becomes delocalized. 
At the same time as $\lambda$ becomes large, the magnitude of the 
energy $E_0\propto 1/\lambda^2$ decreases. For $\lambda\to\infty$,
we recover the free SE as the potential $U(x)\to 0$ in 
Eq.~(\ref{Sec:cosh-variant-a}) and also the nonlinear term 
$F\bigl[\Psi^\ast\Psi\bigr]\to0$ in Eq.~(\ref{Sec:cosh-variant-e}). 
If we apply a Galilean boost according to (\ref{Eq:soliton-in-motion})
and take $\lambda\to\infty$, the solution is of course 
not normalizable and corresponds to a plane wave.

The opposite limit of small $\lambda$ is also interesting. 
The potential of the SE becomes deeper and $E_0$ becomes more negative.
In the NSE, the magnitude of the nonlinear term increases 
(it becomes more negative) since $A$ is proportional 
to $1/\lambda^2$ in Eq.~(\ref{Sec:cosh-variant-e}) and the 
soliton becomes more strongly localized.
This picture remains correct for arbitrarily small, but non-zero 
$\lambda$. In the strict limit $\lambda\to 0$ the potential of the 
SE (and the nonlinear term of the NSE) become singular, the ground 
state energy $E_0\to-\,\infty$, while the ground state wave function becomes 
strongly localized and approaches $|\psi_0(x,t)|^2\to \delta(x)$.
In Appendix~\ref{App:A}, we show that despite this extreme localization 
of the state for $\lambda\to0$, Heisenberg's uncertainty principle is 
always valid.  

For completeness, we remark that the solution $\phi_0(r)=1/\cosh(r/a)^{1/\lambda}$
exists also in $N\ge 2$ dimensions for a generalized potential and a 
generalized nonlinearity which then both have additional structures 
proportional to $(N-1)$. The situation is similar to the case discussed in 
Sec.~\ref{Sec-6:one-over-cosh-N}, and we refrain from showing the results.

\newpage
\section{\boldmath Example of a NSE from piecewise potential}
\label{Sec-8:tan2}

Some exactly solvable quantum problems are given in terms of potentials 
which are defined piecewise. Our next example is of this type. We will 
see that it is possible to derive an NSE also in such a case. In a 
one-dimensional quantum system, we consider the potential given by
\be\label{Sec:tan2-a}
        U(x) = \frac{\hbar^2}{2 m L^2}\, 
        \beta(\beta-1) \tan ^2\left(\frac{x}{L}\right)
        \quad \mbox{for} \quad |x| < \frac{\pi\,L}{2} \;, 
\ee 
and infinite for $|x| \ge \pi\,L/2$ where $L$ is a positive parameter of 
dimension length and $\beta>1$ is dimensionless. The ground state solution of 
the SE with the potential (\ref{Sec:tan2-a}) is for $|x| < \pi\,L/2$ given by
\be\label{Sec:tan2-b}
      \Psi_0(t,x) = c_0\,\phi_0(x)\,e^{-iE_0t/\hbar},
      \quad \phi_0(x) = \biggl(\cos \frac{x}{L}\biggr)^\beta, 
      \quad c_0 = 
      \sqrt{\frac{\beta\,\Gamma(\beta)}{\sqrt{\pi}\,\Gamma(\beta+\tfrac12)L}}\,,
      \quad E_0 = \frac{\hbar^2\beta}{2mL^2} 
\ee
and zero elsewhere. The wave function can be inverted and used to express
the potential in terms of $\phi_0(x)$ as follows
\be\label{Sec:tan2-c}
      U(x) = 
      \frac{\hbar^2}{2 m L^2}\,\beta(\beta -1)\left(\phi_0(x)^{-2/\beta }-1\right)
      \quad \mbox{for} \quad |x| < \frac{\pi\,L}{2} \;.
\ee 
The resulting analytically solvable  NSE is then given by 
Eq.~(\ref{Eq:NSE}) with the nonlinear term defined as
\be\label{Sec:tan2-e}
      F\bigl[\Psi^\ast\Psi\bigr] = A \,G\bigl(|\Psi/c_0|\bigr)\,,
      \quad G(\phi) = \beta(\beta -1) \left(\phi^{-2/\beta }-1\right) ,
      \quad A = \frac{\hbar ^2}{2 L^2 m}\,.
\ee

\begin{figure}[b!] 
\includegraphics[width=0.32\linewidth]{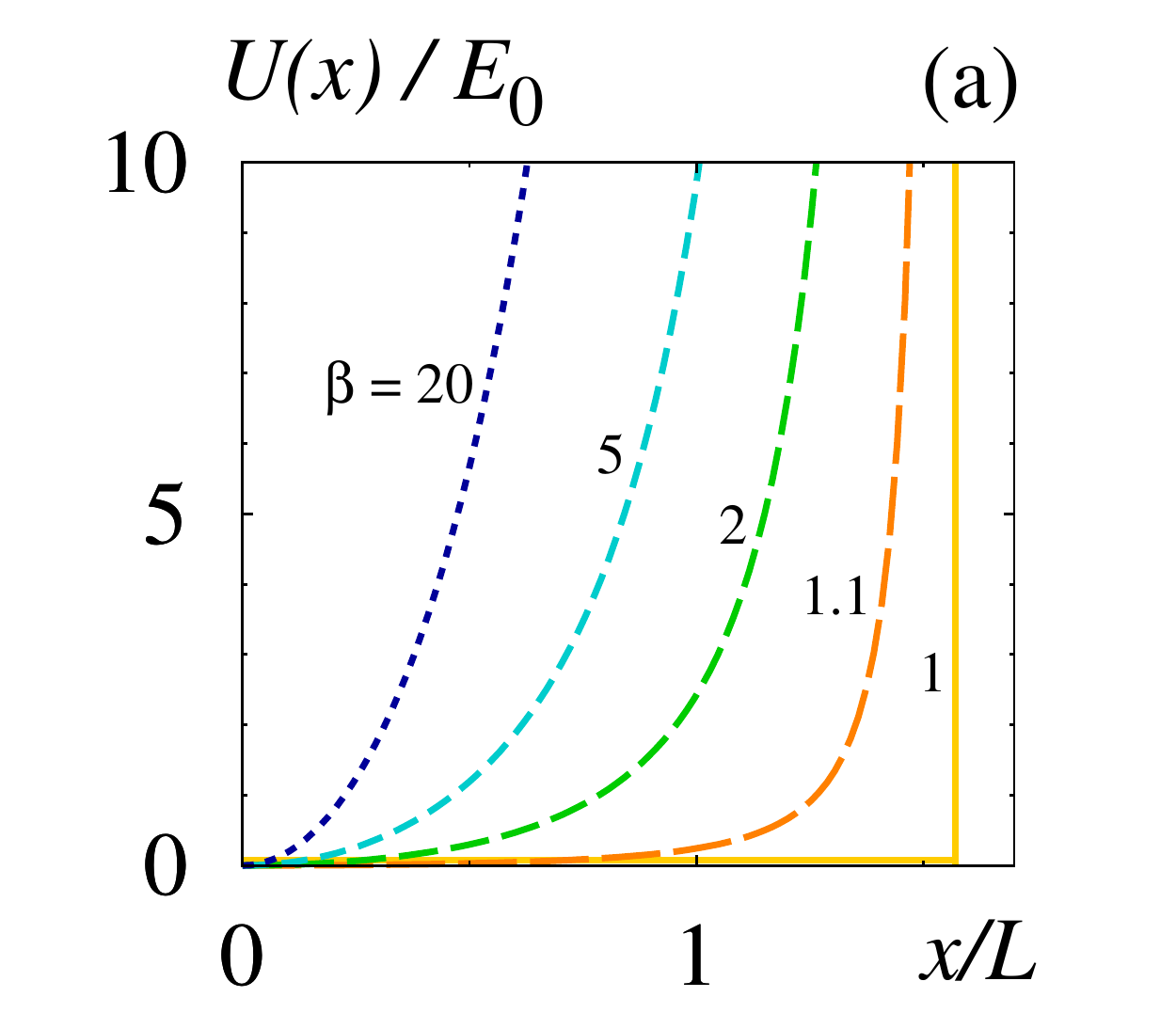}
\includegraphics[width=0.32\linewidth]{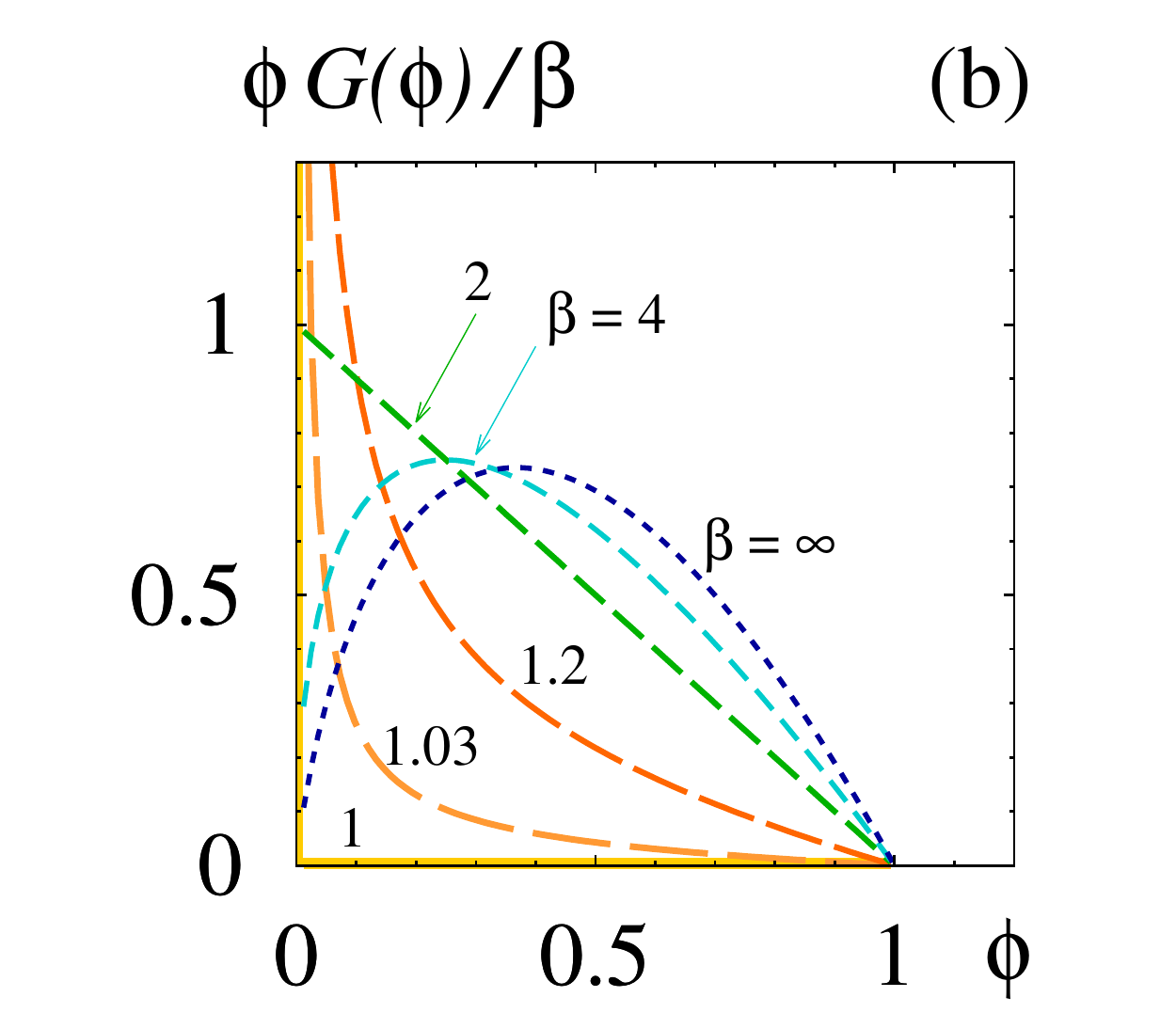}
\includegraphics[width=0.32\linewidth]{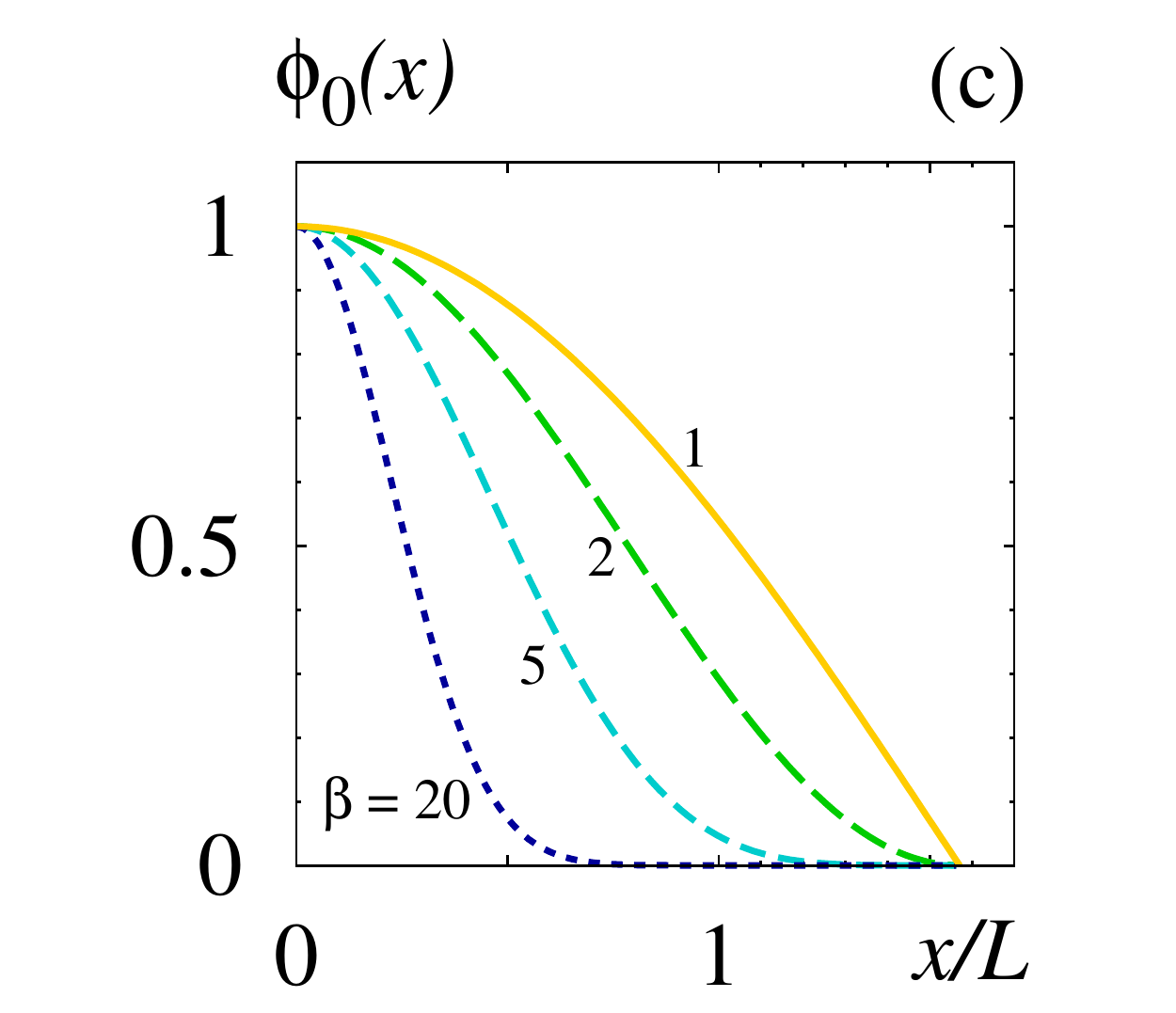}
\caption{
(a) The potential $U(x)$ in Eq.~(\ref{Sec:tan2-a}) in units of $|E_0|$ for 
selected values of $\beta$. The limiting case $\beta=1$ corresponds to the 
familiar square well potential.
(b) The nonlinear term $\phi\, G(\phi)$ with $G(\phi)$ defined in 
Eq.~(\ref{Sec:tan2-e}) normalized with respect to the parameter $\beta$
to better illustrate the scaling in the large-$\beta$ limit. 
(c) The solution $\phi_0(x)=1/\cos(x/L)^\beta$ for selected values of
$\beta$. 
\label{Fig-5:tan2}}
\end{figure}

The potential, nonlinear term, and $\phi_0(x)$ are shown in 
Fig.~\ref{Fig-5:tan2} for selected values of $\beta$. For $\beta\to 1$,
the potential (\ref{Sec:tan2-a}) approaches the familiar infinite square 
well potential, while for $\beta\gg 1$, the potential 
becomes very steep, see Fig.~\ref{Fig-5:tan2}a.
In the limit $\beta\to1$, also the nonlinear function has formally 
a ``square well-type shape'' with the properties (i) $G(\phi)=0$ for 
$\phi\neq0$, and (ii) $G(\phi)\to\infty$ as $\phi\to0$ as illustrated 
in Fig.~\ref{Fig-5:tan2}b. However, although $\beta$ can be infinitesimally 
close to unity, the NSE can only be solved for $\beta>1$. 
In the limit of large $\beta$, the non-linear 
function grows with $\beta$. But when normalized  with respect to $\beta$, 
the nonlinearity has the limit $\lim_{\beta\to\infty}\phi\,G(\phi)/\beta = -2\,\phi\,\ln\phi$, 
as depicted in Fig.~\ref{Fig-5:tan2}b. 
As $\beta\to1$, the solution $\phi_0(x)$ approaches 
the shape $\cos(x/L)$ familiar from the square well potential, while for 
$\beta\gg1$ it becomes strongly localized, see Fig.~\ref{Fig-5:tan2}c. 
In the limit $\beta\to\infty$, the function $\phi_0(x)\to 0$ for $x\neq 0$, 
and the normalized wave function takes the limit 
$\lim_{\beta\to\infty}|\psi_0(x,t)|^2 = \delta(x)$.
Despite the strong localization of the wave function for $\beta\to\infty$,
Heisenberg's uncertainty relation remains valid because the shrinking of
the position uncertainty $\Delta x\to 0$ is accompanied by the corresponding
spread of the momentum uncertainty $\Delta p\to\infty$. Notice also that
$E_0$ diverges as $\beta$ grows. For any $\beta<\infty$ it is always 
$\Delta p\,\Delta x > \frac12\,\hbar$, and 
the uncertainty relation becomes 
an equality in the limit $\beta\to\infty$. The situation is analog to the 
limit $\lambda\to0$ in Sec.~\ref{Sec-7:one-over-cosh-variant} which is
discussed in detail in App.~\ref{App:A}.

\newpage

\section{\boldmath Nonlinear theory with $\delta$-function type 
limiting case}
\label{Sec-9:tam-1D}

In this section, we consider the one-dimensional potential given by
the expression
\be\label{Eq:tam-1D-01}
        U(x,b_0) = -\;\frac{\hbar^2b_0^2}{2 a m}\;
        \left[\frac{1}{\left(x^2+b_0^2\right){ }^{3/2}}
             +\frac{1}{a \left(x^2+b_0^2\right)}\right]\,,
\ee
where $a>0$ and $b_0 > 0$ are constants with the dimension of length.
The ground state solution of the ordinary SE with the potential in 
Eq.~(\ref{Eq:tam-1D-01}) is given by
\be\label{Eq:tam-1D-WF}
     \Psi_0(t,x)=c_0\phi_0(x)\,e^{-iE_0t/\hbar}, 
     \quad  \phi_0(x) = 
     \exp\biggl(\frac{b_0}{a}-\frac{\sqrt{x^2+b_0^2}}{a}\biggr), 
     \quad  E_0=-\,\frac{\hbar^2}{2a^2m}\,.
\ee
We could not find an analytic expression for the normalization constant 
$c_0$ valid in the general case, though it can be computed numerically 
if needed. The expression for $c_0$ is not of importance for the following. 
Notice that $\phi_0(0)=1$ in accordance with Eq.~(\ref{Eq:norm-phi-0}).

The radial function $\phi_0(x)$ can be inverted and the potential expressed as
\be\label{Eq:tam-1D-NSE-0}
    U(x,b_0) = \frac{\hbar ^2}{2 a^2 m}\,G\bigl(\phi_0(x),b_0\bigr),
    \quad G(\phi,b_0) = 
    \frac{b_0^2}{a^2}\,\frac{1-\ln (e^{-b_0/a}\phi)}{\ln^3(e^{-b_0/a}\phi)}
    \,.
\ee
In this way, we obtain the exactly solvable NSE in Eq.~(\ref{Eq:NSE})
with the nonlinearity 
\be\label{Eq:tam-1D-NSE-1}
      F\bigl[\Psi^\ast_{ }\Psi\bigr]
      = A\,G\bigl(|\Psi|/c_0,b_0\bigr)\,,
      \quad A = \frac{\hbar^2}{2m\,a^2}\,,
\ee 
with $G(\phi,b_0)$ defined in Eq.~(\ref{Eq:tam-1D-NSE-0}).
The nonlinear theory (\ref{Eq:tam-1D-NSE-1}) has the analytic 
solution (\ref{Eq:tam-1D-WF}) and can describe traveling 
solutions according to (\ref{Eq:soliton-in-motion}).

We defined the potential in Eq.~(\ref{Eq:tam-1D-01}) for $b_0>0$ and 
excluded the case $b_0=0$. But the limit $b_0\to 0$ can be taken, and it 
is indeed very interesting. In this limit, the potential 
(\ref{Eq:tam-1D-01}) has the properties
\ba\label{Eq:tam-1D-lim-2}
    \mbox{(i)} && \;\;\lim_{b_0\to0} U(x,b_0) \;=\; 0 
    \quad {\rm for} \quad  x\neq0\,, 
    \nonumber\\
    \mbox{(ii)}&& \int\limits_{-\infty}^\infty\!\!{\rm d}x\; U(x,b_0) 
    \;=\; -\frac{\hbar^2(2 a+\pi b_0)}{2 a^2 m}
    \quad {\rm for} \quad  b_0\neq0\,.
\ea
These are basically the defining equations for a $\delta$-function,
i.e.\ in the limit that $b_0\to0$, the potential reduces to the
$\delta$-function potential
\be\label{Eq:tam-1D-lim-3}
      U(x) \equiv \lim_{b_0\to0} U(x,b_0)
           = - \frac{\hbar^2}{a\,m}\,\delta(x)\,,
\ee
while the wave function (\ref{Eq:tam-1D-WF}) reduces in this 
limit to the known solution of this familiar textbook potential, namely
\be\label{Eq:1D-tam-1D-WF}
     \Psi_0(t,x)=c_0\,\phi_0(r)\,e^{-iE_0t/\hbar}, 
     \quad  \phi_0(x) = 
     \exp\biggl(-\frac{|x|}{a}\biggr), 
     \quad  c_0=\frac{1}{\sqrt{a}}\,,
     \quad  E_0=-\,\frac{\hbar^2}{2a^2m}\,.
\ee

The formulation of the pertinent NSE must be performed with similar care. 
The nonlinear function (\ref{Eq:tam-1D-NSE-0}) satisfies
\ba\label{Eq:1D-tam-1D-NSE-1}
     {\rm (i)} && \lim_{b_0\to0}  G(\phi,b_0) = 0 \quad
     {\rm for} \quad  0\le \phi < 1,\nonumber\\
     {\rm (ii)} && \int\limits_0^1{\rm d}\phi \; G(\phi,b_0) =
     -\,\frac12 - \frac{b_0}{2a} - 
     \frac{b_0^2}{2a^2} \,e^{b_0/a}\, \text{Ei}(-b_0/a) \quad
     {\rm for} \quad  b_0 > 0,
\ea
where ${\rm Ei}(-y) = \int_y^\infty du\,e^{-u}/u$ denotes the exponential 
integral. The properties in Eq.~(\ref{Eq:1D-tam-1D-NSE-1}) define in the
limit $b_0\to0$ a $\delta$-function, this time with support at $\phi=1$, i.e.\
\be\label{Eq:1D-tam-1D-NSE-2}
     \lim_{b_0\to0}  G(\phi,b_0) = -\,\delta(1-\phi)\,,
\ee
with the convention that integrating a $\delta$-function up to a limit
which coincides with its support yields $\int_0^c{\rm d}u\,\delta(u-c)=\frac12$
for $c>0$.
In this way, we find an unusual exactly solvable NSE, namely
\be\label{Eq:tam-1D-NSE-3}
      i\hbar \,\frac{\partial\Psi}{\partial t} 
      = -\,\frac{\hbar^2}{2m}\,\bigtriangleup\Psi 
      - A\,\delta\bigl(1-|\Psi|/\sqrt{a}\bigr)\,,
      \quad A = \frac{\hbar^2}{2m\,a^2}\,.
\ee 
The nonlinearity in this problem is nonzero only when the spatial part 
of the wave function $\phi(x)$ becomes unity. This is the case for the 
solution $\phi_0(x)$ in Eq.~(\ref{Eq:1D-tam-1D-WF})
(cf.\ Eq.~(\ref{Eq:norm-phi-0}) for conventions) only at $x=0$ corresponding 
to the only point where the limiting potential (\ref{Eq:tam-1D-lim-3}) is 
nonzero. The very presence of the singular nonlinearity in 
Eq.~(\ref{Eq:tam-1D-NSE-3}) can be verified only by integrating the 
(time-independent version of the) NSE in Eq.~(\ref{Eq:tam-1D-NSE-3})
over an infinitesimal interval $[-\epsilon,\,\epsilon]$ enclosing the
point $x=0$, i.e.\ in very much the same way the singular potential 
(\ref{Eq:tam-1D-lim-3}) is treated in the ordinary SE.

\begin{figure}[t!] 
\includegraphics[width=0.32\linewidth]{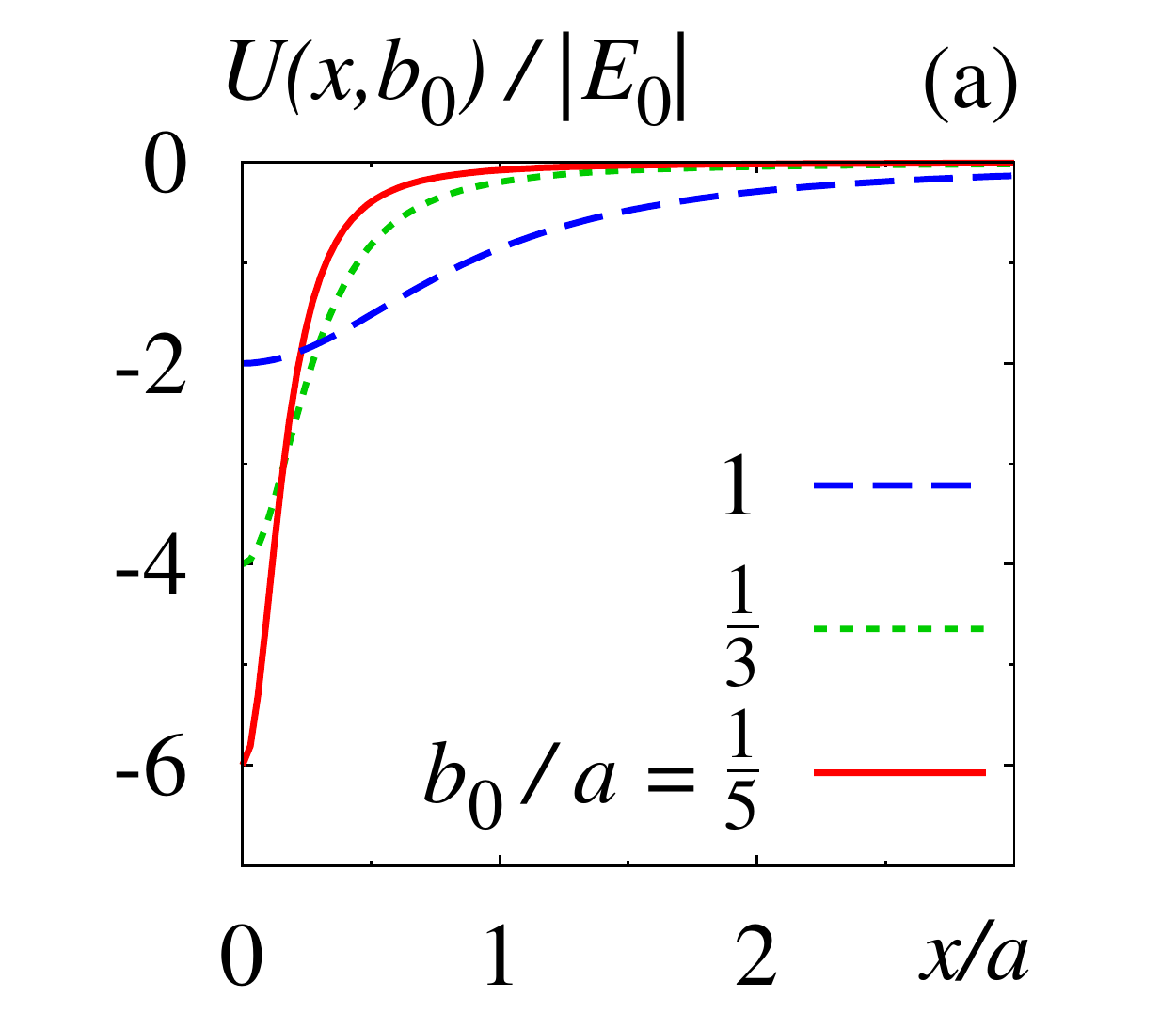}
\includegraphics[width=0.32\linewidth]{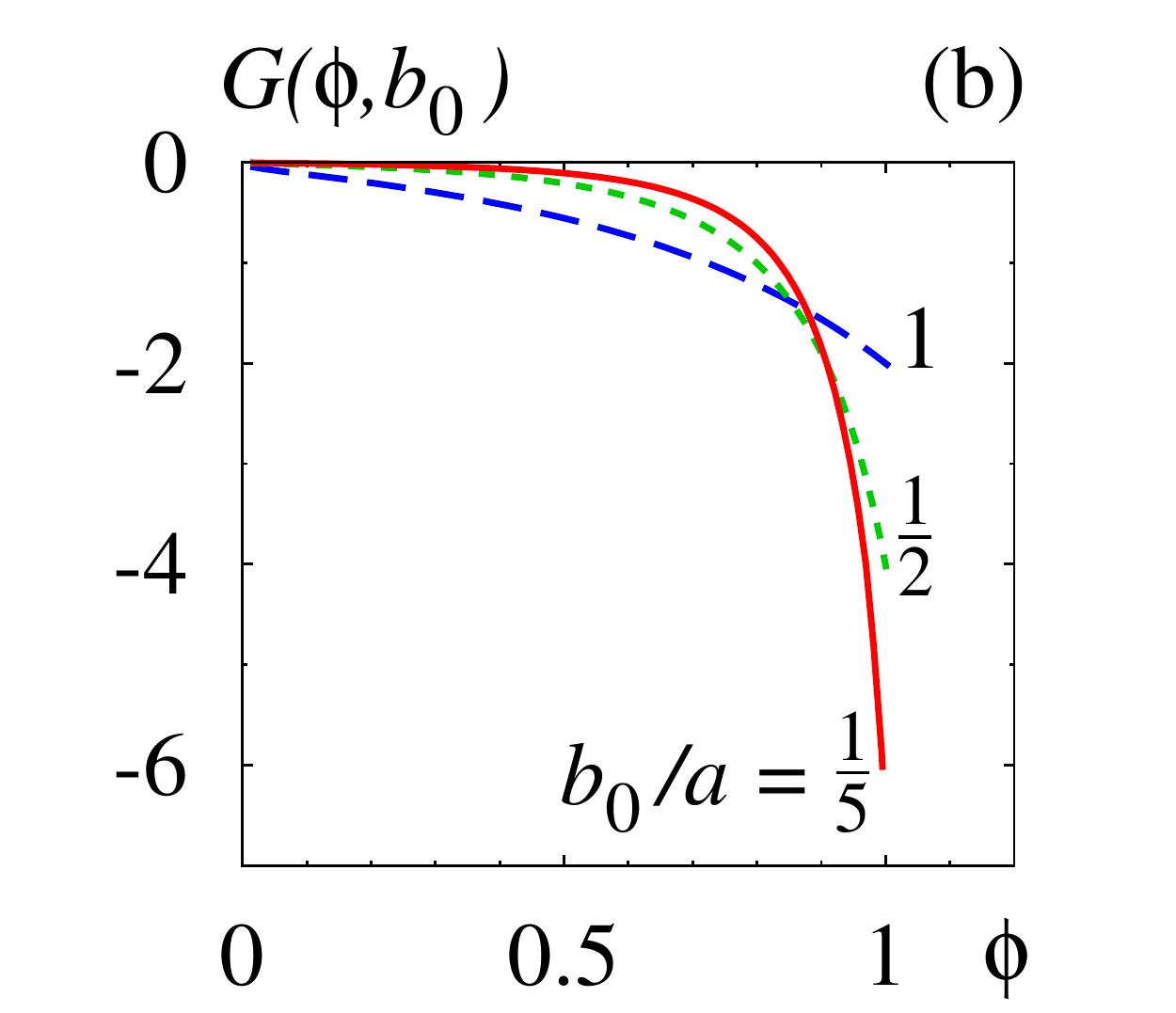}
\includegraphics[width=0.32\linewidth]{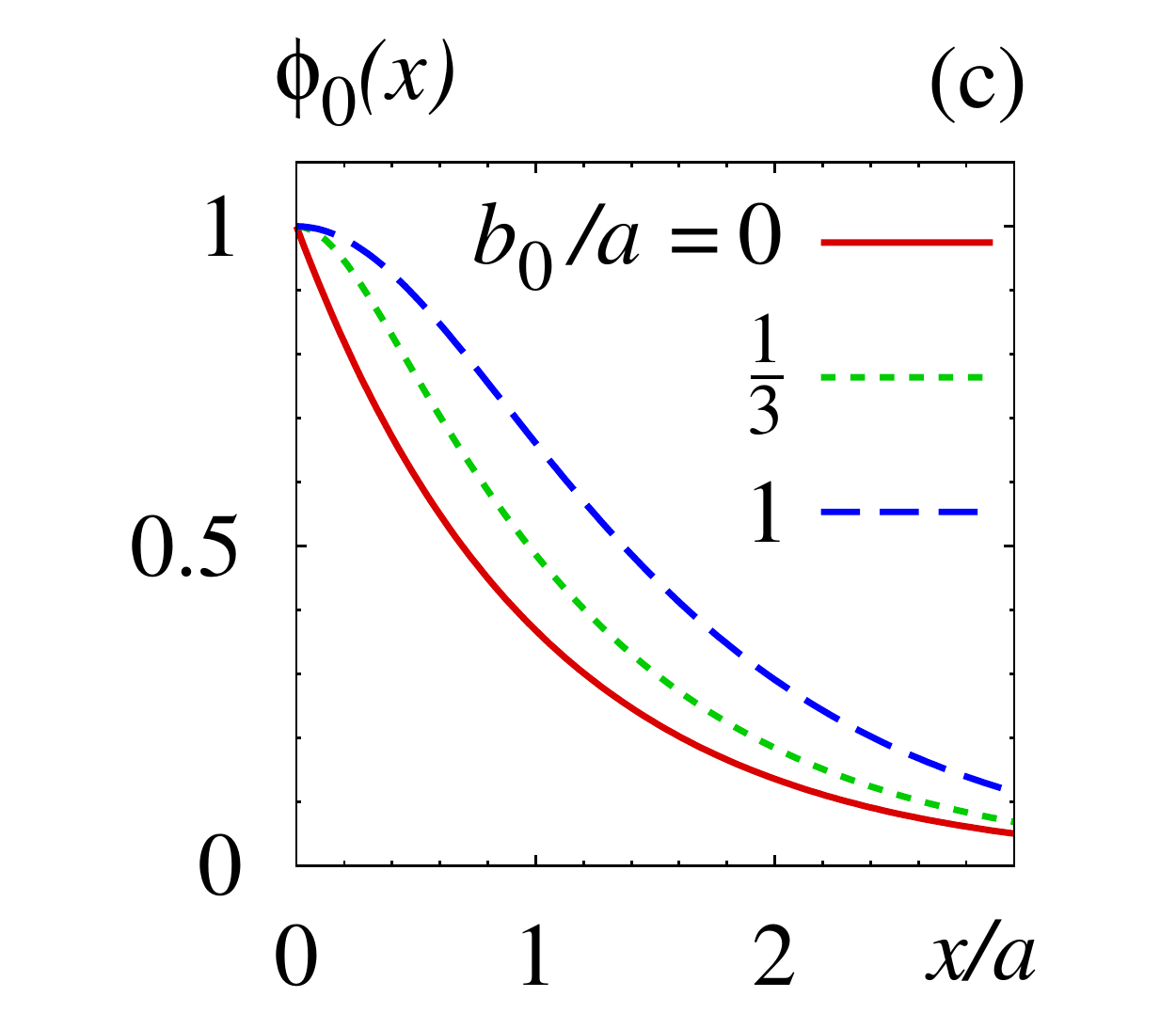}
\caption{
(a) The potential $U(x,b_0)$ in Eq.~(\ref{Eq:tam-1D-01}) in units of 
$|E_0|$ for selected values of $b_0$. For $b_0\to 0$ the potential 
reduces to an attractive $\delta(x)$ potential.
(b) The nonlinear term $G(\phi,b_0)$ defined in Eq.~(\ref{Eq:tam-1D-NSE-0}) 
which becomes $-\,\delta(1-\phi)$ in the limit $b_0\to0$.
(c) The solution $\phi_0(x)$ for selected values of $b_0$ including
the limit $b_0\to 0$.
\label{Fig-6:tam-1D}}
\end{figure}

In Fig.~\ref{Fig-6:tam-1D}, we show the potential, nonlinear term and
spatial part $\phi_0(x)$ for selected values of $b_0$. As the parameter
$b_0$ decreases, the potential and nonlinear term become narrower and
deeper as shown in Figs.~\ref{Fig-6:tam-1D}a and \ref{Fig-6:tam-1D}b. 
The minimum of the potential in units of $|E_0|$ and the non-linear 
function are given by $U(0,b_0)/|E_0|=G(1,b_0)=-a/b_0-1$ and go to 
minus infinity for $b_0\to0$. 
Both functions eventually approach the corresponding singular limits 
in Eqs.~(\ref{Eq:tam-1D-lim-3},~\ref{Eq:1D-tam-1D-NSE-2}) which cannot 
be depicted. The wave function $\phi_0(x)$ is regular in the limiting 
case $b_0\to0$ and shown in Fig.~\ref{Fig-6:tam-1D}c.

In the opposite limit $b_0\to\infty$, the potential $U(x,b_0)\to E_0$ 
becomes a trivial constant, the non-linearity $G(\phi,b_0)\to 0$.
After a Galilean boost (\ref{Eq:soliton-in-motion}), the wave-function 
describes a non-renormalizable plane wave solution. In other words,
we recover a free SE in the limit $b_0\to\infty$.

{}

\newpage
\section{Three-dimensional nonlinear theory from Coulomb potential}
\label{Sec-10:Coulomb}

As our final example, we choose another well-familiar analytically
solvable quantum mechanical potential, namely the Coulomb potential 
in $N=3$ dimensions. The potential is given by
\be\label{Eq:Coulomb-01}
        U(r) = -\,\frac{e^2}{4\pi\varepsilon_0} \,\frac1r 
             = - \frac{\hbar^2}{a_B\,m}\,\frac1r
\ee
where $a_B= 4\pi\varepsilon_0 \hbar^2 / (e^2m) = \hbar/(\alpha mc)$ denotes 
the Bohr radius, $m$ the reduced mass, and $\alpha$ the fine structure
constant. The ground state energy and wave function are given by
\be\label{Eq:Coulomb-WF}
     \Psi_0(t,\vec{x})=c_0\,\phi_0(r)\,e^{-iE_0t/\hbar}, 
     \quad  \phi_0(r) = \exp\biggl(-\frac{r}{a_B}\biggr), 
     \quad  c_0 = \biggl(\frac{1}{\pi a^3_B}\biggr)^{\!1/2}\,,   
     \quad  E_0=-\,\frac{\hbar^2}{2a^2_B m} = -\frac12\,\alpha^2\, mc^2\,.
\ee
This wave function can be inverted such that we obtain
\be
    r = -\frac{a_B}{2}\;
    \ln\biggl(\frac{|\Psi_0(t,\vec{x})|^2}{|c_0|^2}\biggr)\,.
\ee
Hence, we can rewrite the Coulomb potential as
\be
    U(r) = \frac{2\hbar^2}{m\,a^2_B}\,
           \frac{1}{\ln\bigl(|\Psi_0(t,\vec{x})|^2/|c_0|^2\bigr)}\,.
\ee
In this way, we find the exactly solvable NSE in Eq.~(\ref{Eq:NSE})
where the nonlinear term $F[\Psi^\ast\Psi]$ is given by
\be\label{Eq:Coulomb-NSE-2}
      F\bigl[\Psi^\ast_{ }\Psi\bigr]
      = \frac{A}{\ln\bigl(B \,\Psi^\ast_{ }\Psi\bigr)}
      = A\,G\bigl(|\Psi/c_0|\bigr)\,,
      \quad A = \frac{2\hbar^2}{m\,a^2_B},
      \quad B = \pi a^3_B ,
      \quad G(\phi) = \frac{1}{\ln(\phi^2)}\,.
\ee 
The nonlinear theory (\ref{Eq:NSE},~\ref{Eq:Coulomb-NSE-2}) has 
the analytic solution (\ref{Eq:Coulomb-WF}) and can describe traveling 
solitons according to Eq.~(\ref{Eq:soliton-in-motion}). To the best of our 
knowledge, this nonlinear theory has not been discussed in literature before.

\begin{figure}[b!] 
\includegraphics[width=0.32\linewidth]{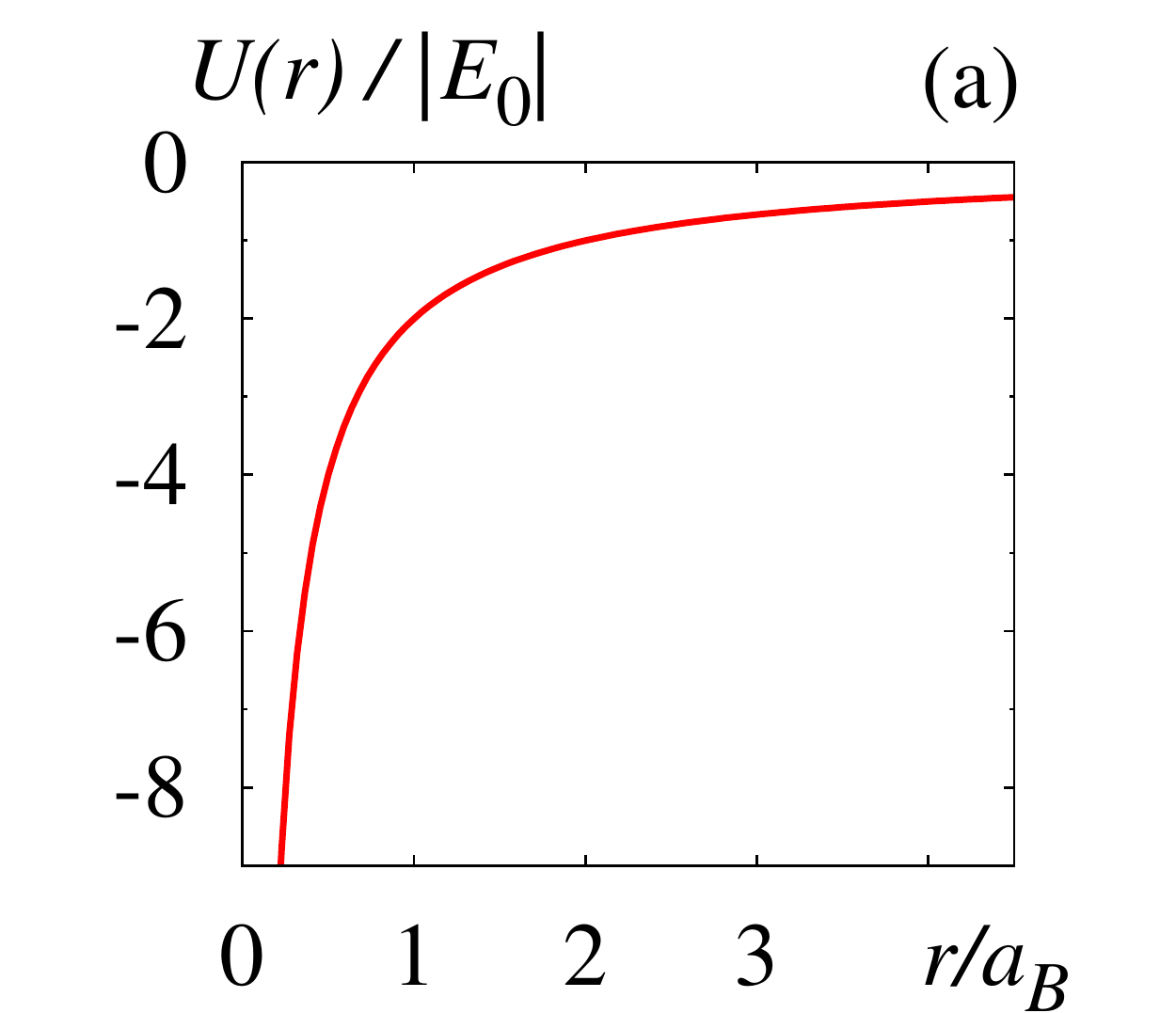}
\includegraphics[width=0.32\linewidth]{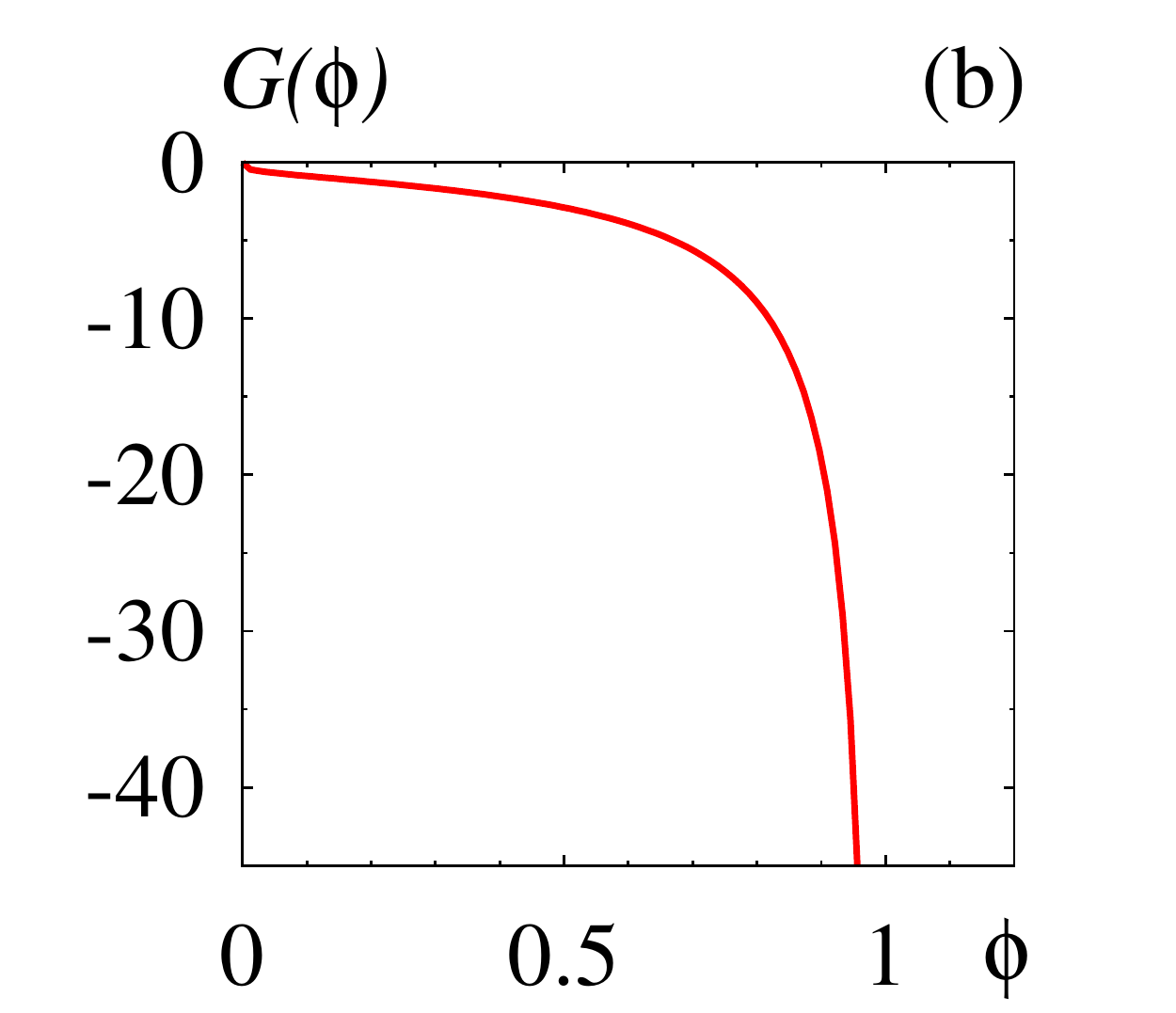}
\includegraphics[width=0.32\linewidth]{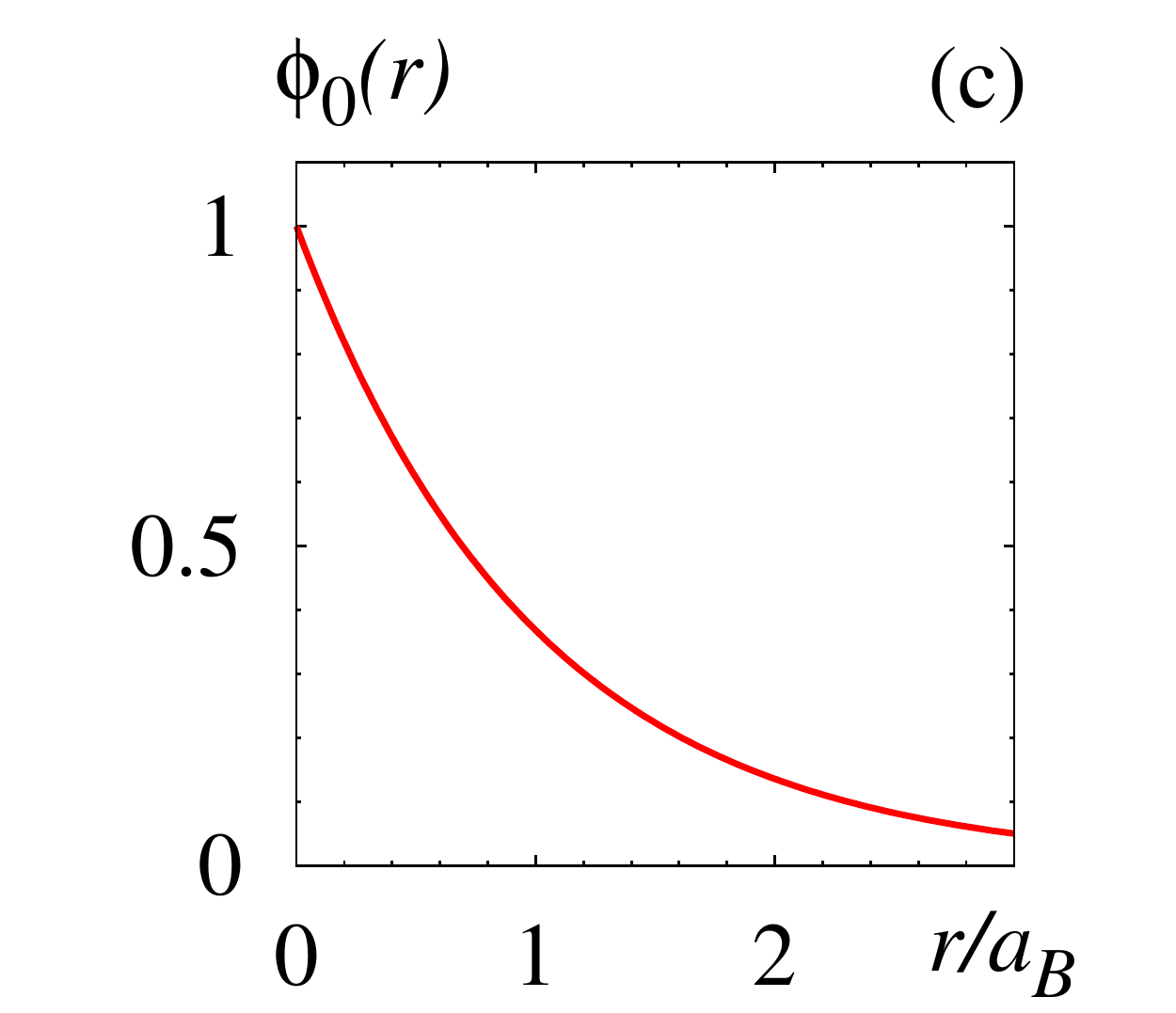}
\caption{(a) The three-dimensional Coulomb potential in units of 
$|E_0|$ as function of $r$ in units of the Bohr radius $a_B$.
(b) The nonlinear term $G(\phi)$ of the NSE defined in 
Eq.~(\ref{Eq:Coulomb-NSE-2}).
(c)~The radial part $\phi_0(r)$ of respectively the Coulomb ground 
state wave function or the soliton of the NSE (\ref{Eq:NSE}) with 
the nonlinearity (\ref{Eq:Coulomb-NSE-2}).
\label{Fig-2:Coulomb}}
\end{figure}

In Fig.~\ref{Fig-2:Coulomb}, we depict for completeness the Coulomb 
potential $U(r)$, the nonlinear function $G(\phi)$, and the radial 
function $\phi_0(r)$. The function $G(\phi)$ is throughout negative 
for $\phi>0$ and diverges when $\phi\to1$ which is in one-to-one 
correspondence to the divergence of the Coulomb potential at $r\to 0$.

\newpage
\section{Conclusions}
\label{Sec-11:conclusions}

In this work, we have presented a method to construct analytically
solvable nonlinear extensions of the Schr\"odinger equation (NSE)
starting from an ordinary analytically solvable Schr\"odinger
equation (SE). We have illustrated the method through several 
examples in which the potential $U(\vec{x})$ of the SE in Eq.~(\ref{Eq:SE})
was systematically transformed into a nonlinear term $F[\Psi^\ast\Psi]$
of the NSE in Eq.~(\ref{Eq:NSE}). 

Starting from respectively the harmonic potential or (a special case) 
of the Rosen-Morse potential we rederived well-known soliton solutions
of nonlinear theories, namely the Gausson in a general number of space
dimensions $N$ and the one-dimensional 1/cosh soliton 
\cite{BialynickiBirula:1976zp,BialynickiBirula:1979dp,Oficjalski:1978,
Zakharov-Shabat}. In several other cases, we have derived exact soliton
solutions of non-linear theories which, to the best of our knowledge,
have not been discussed previously in literature. This includes among 
others a nonlinear theory derived from the SE with the Coulomb potential 
in $N=3$ dimensions. Another interesting example was a regular one-dimensional 
potential which can be transformed into the attractive $\delta(x)$ potential by 
taking one of the parameters of this potential to approach a specific limit.
The regular potential as well as the singular $\delta(x)$ potential can 
both be used to construct exactly solvable NSE with interesting soliton
solutions.

The quantum mechanical potentials explored in this work have in common
that they are symmetric, i.e.\ $U(\vec{x})=U(r)$ with $r=|\vec{x}|$ in 
$N>1$ space dimensions or $U(x)=U(|x|)$ in $N=1$ space dimensions. 
Another common feature is that the considered potentials have a 
single minimum which can be finite or infinite.
It is an interesting question whether the method can be generalized to 
construct exactly solvable nonlinear soliton theories also under more 
general conditions, e.g.\ starting from non-symmetric
potentials or from double-well type potentials.

Another interesting future direction could be to explore systematically
methods like Lie algebra techniques and self-similar potentials 
\cite{Shabat:1992,Spiridonov:1992} or the more general concept of shape 
invariant potentials \cite{Gendenshtein:1983skv} and other supersymmetric 
methods in quantum mechanics \cite{Cooper:1994eh,Bougie:2012} or whether
novel soliton solutions can be found in non-hermitian PT symmetric quantum 
systems in analogous ways \cite{Ahmed:2001gz,Musslimani:2008zz,Konotop:2016eny}.
These interesting questions will be addressed in future studies.

\ \\
\noindent{\bf Acknowledgments.} 
This work was supported by the National Science Foundation 
under the Contract No.\ 1812423 and 2111490.
This work was supported in part also by the Department of 
Energy within framework of the QGT Topical Collaboration

\appendix

\section{Heisenberg's uncertainty principle for extremely 
localized wave functions}
\label{App:A}

\begin{figure}[t!] 
\includegraphics[width=0.32\linewidth]{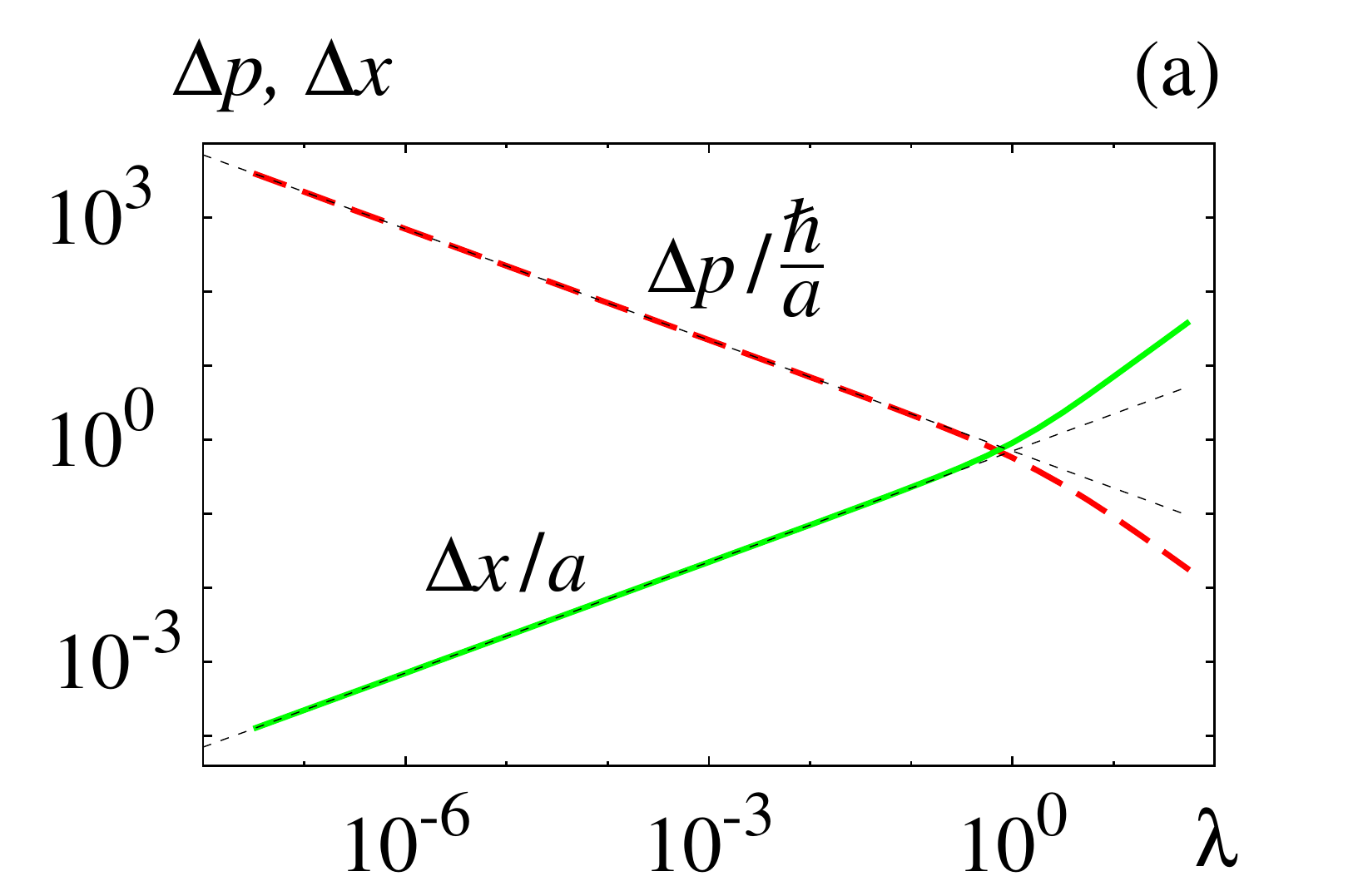} \hspace{1cm}
\includegraphics[width=0.32\linewidth]{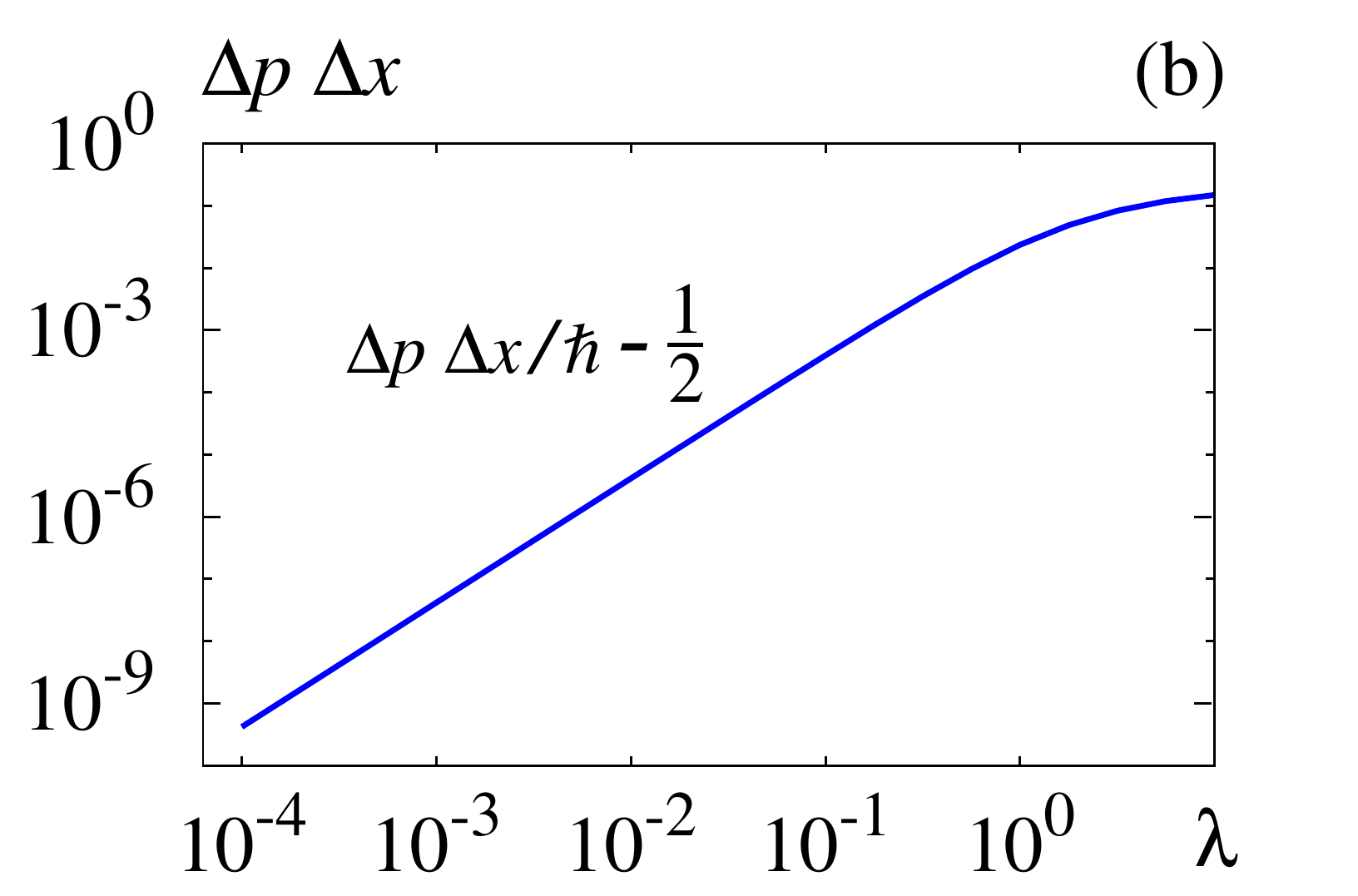}
\caption{(a) 
    Position uncertainty $\Delta x$ in units of $a$ and 
    momentum uncertainty $\Delta p$ in units of $\hbar/a$ 
    in the ground state (\ref{Sec:cosh-variant-b}) discussed in 
    Sec.~\ref{Sec-7:one-over-cosh-variant} as functions of $\lambda$. 
    The dotted lines show respectively the asymptotic results for
    $\Delta p$ and $\Delta x$ in Eq.~(\ref{Eq-App:cosh-localized-3}).
    (b) 
    Test of Heisenberg's uncertainty relation expressed as
    $\Delta x\,\Delta p/\hbar - \frac12$ which must be positive 
    as is the case, see text.
    \label{Fig-7:Heisenberg-uncertainty}}
\end{figure}

In Sec.~\ref{Sec-7:one-over-cosh-variant}, we discussed the exactly solvable 
quantum potential (\ref{Sec:cosh-variant-a}). In this Appendix, we investigate 
in detail the limit $\lambda\to0$ in which the ground state wave function 
(\ref{Sec:cosh-variant-b}) behaves such that the probability density has the 
properties 
\ba\label{Eq-App:cosh-localized-1}
      (i) &&  \lim_{\lambda\to0}|\Psi_0(t,x)|^2 
              = 0 
              \quad \mbox{for} \quad x\neq0 \, , \nonumber \\
      (ii) && \int\limits_{-\infty}^\infty\!\! dx\;|\Psi_0(t,x)|^2 = 1
              \quad \mbox{for} \quad \lambda \neq0 \, .
\ea
These properties imply that the probability density becomes extremely 
localized as 
\be\label{Eq-App:cosh-localized-2}
      \lim_{\lambda\to0}|\Psi_0(t,x)|^2  = \delta(x), 
\ee
and exhibits an obviously vanishing position uncertainty 
$\Delta x$.
It is interesting to ask whether such an extremely localized state 
satisfies Heisenberg's uncertainty principle. For $\lambda\neq 0$ it 
is always $\Delta p\;\Delta x > \frac{\hbar}{2}$. We refrain from showing 
the bulky analytic expressions for $\Delta p$ and $\Delta x$ which,
if needed, can be found easily with {\tt mathematica} and are given 
in terms of Gamma functions and hypergeometric functions.
The results for $\Delta x$ and $\Delta p$ are shown in 
Fig.~\ref{Fig-7:Heisenberg-uncertainty}.
As $\lambda$ decreases, the position uncertainty $\Delta x$ becomes 
smaller while the momentum uncertainty $\Delta p$ increases. For 
infinitesimally small (but non-zero) $\lambda$, the uncertainties 
behave as
\ba\label{Eq-App:cosh-localized-3}
      \Delta x = a \sqrt{\frac{\lambda}{2}} + \dots\;, \quad
      \Delta p = \frac{\hbar}{a}\;\frac{1}{\sqrt{2\lambda}} + \dots\;, 
\ea
where the dots indicate positive higher order corrections such 
that $\Delta x\,\Delta p > \frac{\hbar}{2}$ for all $\lambda>0$. 
The leading terms in Eq.~(\ref{Eq-App:cosh-localized-3}) approximate 
the momentum and position uncertainties to within ${\cal O}(2\,\%)$ 
already for $\lambda\lesssim 10^{-1}$ and describe $\Delta p$ and 
$\Delta x$ over several orders of magnitude for $\lambda\ll 1$ in 
Fig.~\ref{Fig-7:Heisenberg-uncertainty}.
From Eq.~(\ref{Eq-App:cosh-localized-2}), we see that 
$\lim_{\lambda\to0} \Delta p \;\Delta x = \frac{\hbar}{2}$.
Thus, Heisenberg's uncertainty relation is manifestly valid for 
any value of $\lambda$ including the limit $\lambda\to0$.

The increasing of the momentum uncertainty $\Delta p$ as $\lambda$
becomes small implies a large kinetic energy. In fact, since in this 
stationary state $\la p\ra = 0$, we have 
\be\label{Eq-App:cosh-localized-4}
      \la E_{\rm kin}\ra 
      = \frac{\la p^2\ra}{2m} 
      = \frac{\Delta p^2}{2m} 
      = \frac{\hbar^2}{2ma^2}\;\frac{1}{2\lambda} + \dots\;,
\ee
modulo subleading corrections for $\lambda\ll 1$. Thus, the expectation
value of the kinetic energy diverges as $1/\lambda$ for $\lambda\to0$
which is a consequence of the strong localization of the quantum state.
However, it is important to keep in mind that the total (negative) binding 
energy $E_0$ is proportional to $1/\lambda^2$ in Eq.~(\ref{Sec:cosh-variant-b}). 
Thus, when ``measured in units'' of the absolute value of the total energy, 
the expectation value of the kinetic energy actually behaves as
\be\label{Eq-App:cosh-localized-5}
      \frac{\la E_{\rm kin}\ra}{|E_0|} =
      \frac{\lambda}{2} + \dots\; 
\ee 
with the dots denoting higher order terms. In other words, $\la E_{\rm kin}\ra$ 
becomes negligibly small in the limit $\lambda\to0$ in comparison to 
the total binding energy. This is because the ground state energy is 
dominated by the expectation value of the potential energy with the 
potential (\ref{Sec:cosh-variant-a}) behaving as $U(x)\propto 1/\lambda^2$.
These properties make sense physically for arbitrarily small but non-zero
values of $\lambda$. We deal with a deeply bound and strongly localized state. 
The fact that $\la E_{\rm kin}\ra \ll |E_0|$ means the motion of the particle
becomes negligible as the particle becomes localized due to the strong coupling.
Nevertheless, the uncertainty principle remains valid for any $\lambda$.

It is interesting to notice that the numerical computation of 
$\Delta p$ and $\Delta x$ can be carried out down to much smaller values 
of $\lambda$ than the numerical test of the uncertainty relation.
The reason for that is as follows. At $\lambda=10^{-4}$ the asymptotic 
expressions in Eq.~(\ref{Eq-App:cosh-localized-3}) underestimate 
$\Delta p$ in relative units by about ${\cal O}(2\times10^{-5})$ and 
overestimate $\Delta x$ by about the same amount. We can go down to 
$\lambda = 10^{-8}$ before hitting numerical accuracy limitations for 
$\Delta p$ and $\Delta x$ on the scale
of Fig.~\ref{Fig-7:Heisenberg-uncertainty}a.
However, the over- and underestimates in $\Delta p$ and $\Delta x$ largely compensate each other in the product such that at $\lambda=10^{-4}$ we reach
with $\Delta p\,\Delta x/\hbar-\frac12 = {\cal O}(4\times10^{-10})$ 
our numerical accuracy in 
Fig.~\ref{Fig-7:Heisenberg-uncertainty}b.

The features that (i) ground state energy $E_0\to-\,\infty$ and (ii) 
probability density $|\psi_0(x,t)|^2\to \delta(x)$ occur also in 
the case of the attractive one-dimensional $1/|x|$-potential 
when the $1/|x|$ singularity is ``regulated'' as $1/(|x|+\epsilon)$
and the limit $\epsilon\to0$ is taken \cite{Loudon:1959}.
There is no deeper analogy between our case and the regularized
$1/|x|$ potential. This is rather a generic feature of systems with 
strongly localized and deeply bound ground states. As long as 
the parameter ($\lambda$ in our case or $\epsilon$ in the 
regulated $1/|x|$ potential) is infinitesimally small but 
non-zero, one deals with a well-behaved quantum state.
It has however been questioned whether the strict limit itself of 
such a strongly localized state with $|\psi_0(x,t)|^2\to \delta(x)$ 
constitutes a physical state, see \cite{Andrews:1966}.

\end{document}